\documentclass[fleqn,usenatbib]{mnras}

\usepackage{newtxtext,newtxmath,xcolor}

\usepackage[T1]{fontenc}

\DeclareRobustCommand{\VAN}[3]{#2}
\let\VANthebibliography\thebibliography
\def\thebibliography{\DeclareRobustCommand{\VAN}[3]{##3}\VANthebibliography}

\usepackage{graphicx}	
\usepackage{amsmath}	
\usepackage{stfloats}
\graphicspath{{figure/}}

\newcommand{\msun}{\mbox{${\rm M}_{\odot}$}}

\newcommand{\ct}{{\sc consistent trees}}
\newcommand{\rockstar}{{\sc rockstar}}
\newcommand{\subfind}{{\sc subfind}}

\def\lesssim{\lower.5ex\hbox{$\; \buildrel < \over \sim \;$}}
\def\gtrsim{\lower.5ex\hbox{$\; \buildrel > \over \sim \;$}}

\title[The relationship between galaxy size and halo properties]{The relationship between galaxy size and halo properties: Insights from the IllustrisTNG simulations and differential clustering}

\author[Somerville et al.]{
Rachel S. Somerville$^{1}$\thanks{rsomerville@flatironinstitute.org},
Austen Gabrielpillai$^{2,3,4,5}$,
Boryana Hadzhiyska$^{6,7}$,
Shy Genel$^{1,8}$
\\
$^{1}$Center for Computational Astrophysics, Flatiron Institute, 162 5th Ave, New York, NY 10010, USA\\
$^{2}$Institute for Astrophysics and Computational Sciences, Catholic University of America, USA\\
$^{3}$Astrophysics Science Division, NASA/GSFC, 8800 Greenbelt Rd, Greenbelt, MD 20771, USA\\
$^{4}$Center for Research and Exploration in Space Science and Technology, NASA/GSFC, 8800 Greenbelt Rd, Greenbelt, MD 20771, USA\\
$^{5}$Department of Astrophysics, The Graduate Center, City University of New York, 365 5th Ave, New York, NY 10016, USA\\
$^{6}$Physics Division, Lawrence Berkeley National Laboratory, Berkeley, CA 94720, USA\\
$^{7}$Berkeley Center for Cosmological Physics, Department of Physics, University of California, Berkeley, CA 94720, USA\\
$^{8}$Columbia Astrophysics Laboratory, Columbia University, 550 West 120th Street, New York, NY 10027, USA\\
}

\date{Accepted XXX. Received YYY; in original form ZZZ}

\pubyear{2022}

\begin{document}
\label{firstpage}
\pagerange{\pageref{firstpage}--\pageref{lastpage}}
\maketitle

\begin{abstract}
The physical origin of the radial sizes of galaxies and how galaxy sizes are correlated with the properties of their host dark matter halos is an open question in galaxy formation. In observations, the large-scale clustering of galaxies selected by stellar mass is significantly different for large and small galaxies, and \citet{Behroozi:2022} showed that these results are in tension with some of the correlations between galaxy size and halo properties in the literature. We analyze the IllustrisTNG suite of large volume cosmological hydrodynamic simulations along with dark matter only simulations with matched initial conditions.  We investigate correlations between the ratio of galaxy size to halo virial radius ($r_{\rm gal}/R_{\rm vir}$) and halo spin, concentration, and formation time at redshift $0$--3. We find a significant correlation between $r_{\rm gal}/R_{\rm vir}$ and concentration, but only above a critical value $c \simeq 16$, and we also find a correlation between $r_{\rm gal}/R_{\rm vir}$ and halo formation time. We suggest that galaxy formation history and environment, in addition to halo properties at a given output time, plays an important role in shaping galaxy size. In addition, we directly measure size-based differential clustering in the TNG300 simulation and compare directly with the observational results. We find significant scale-dependent size-based differential clustering in TNG, in qualitative agreement with observations. However, correlations between $r_{\rm gal}/R_{\rm vir}$ and secondary halo properties are not the drivers of the differential clustering in the simulations; instead, we find that most of this signal in TNG arises from satellite galaxies. 
\end{abstract}



\section{Introduction}
\label{sec:introduction}

The radial size of galaxies is observed to be strongly correlated with global galaxy properties, such as the luminosity or mass \citep{Shen:2003,Bernardi:2010,Lange:2015}. This correlation has been observed out to the highest redshifts where galaxy sizes can be robustly measured, although the slope and normalization of the correlation evolves with cosmic time, and depends on galaxy type \citep{Ravindranath:2004,Ferguson:2004,Barden:2005,vanderwel:2014,shibuya:2015, morishita:2024}. 

In the standard $\Lambda$CDM paradigm for galaxy formation \citep{White_Rees:1978,Blumenthal:1985}, galaxies form within dark matter halos, and linking galaxy properties with those of their host halos is the goal of much recent work. One commonly used empirical method for constraining the relationship between dark matter halo properties and galaxy properties is known as \emph{Abundance Matching} (sometimes known as Sub-halo Abundance Matching (SHAM); e.g. \citealp{Moster:2010,Behroozi:2010,Moster:2013,Behroozi:2013}). In this approach, a mapping is sought such that the abundance of galaxies with a given property (such as luminosity or stellar mass) matches that of dark matter halos, as predicted by N-body simulations within an assumed cosmology. One can extend this approach to other galaxy properties, such as the radial size, as follows: first, ``standard'' abundance matching is used to derive the mapping between dark matter halo mass and galaxy stellar mass. The observed size-mass relation can then be used to derive a mapping between halo mass and galaxy size. Since halo mass and halo virial radius have a unique relationship for a particular halo definition and redshift, this results in constraints on the relationship between galaxy size and halo size. Several authors have used this approach to show that, surprisingly, the average ratio between galaxy size and halo size $r_{\rm gal}/R_{\rm halo}$ is close to constant at $z=0$ \citep{Kravtsov:2013,Somerville:2018}. This result applies for both late and early type (star forming and quiescent) galaxies, and the ratio evolves only weakly up to $z\sim 3$ \citep{Somerville:2018,huang:2017}.  

However, there is a significant dispersion in the observed sizes of galaxies at fixed stellar mass \citep{Somerville:2018}. The physical origin of this scatter, and the related question of which higher order halo parameters beyond mass are correlated with $r_{\rm gal}/R_{\rm halo}$, remain open questions. In the classical picture of disk formation \citep{Blumenthal:1986,Dalcanton:1997,Mo:1998,Dutton:2011a}, it is expected that the disk radius should be strongly positively correlated with the halo angular momentum (commonly expressed in terms of the dimensionless spin parameter defined by \citealt{Peebles:1969}). The sizes of early type (spheroidal) galaxies are expected to be driven by mergers, where gas-rich mergers create more compact remnants, and gas-poor mergers lead to more extended remnants \citep{Covington:2008,Hopkins:09a,Covington:2011,Naab:2009,Porter:2014}. 
There have been numerous investigations of the predictions of galaxy sizes in numerical cosmological hydrodynamic simulations \citep{Sales:2009,Genel:2015,Furlong:2017,Genel:2018}. While several relatively recent simulations predict galaxy size-mass relations that are in good agreement with observations both locally and out to $z\sim 3$, it is clear that galaxy sizes in cosmological simulations are quite sensitive to the details of the implementation of physical processes such as stellar and black hole driven winds \citep{Uebler:2014,Agertz:2016}. For example, the sizes at a given stellar mass predicted by the original generation of Illustris simulations were about a factor of two larger than those predicted in the revised IllustrisTNG simulations \citep{Genel:2015,Genel:2018}. These simulations used the same underlying hydrodynamical and gravity codes and initial conditions, and differed primarily in the implementation of stellar winds and Active Galactic Nuclei (AGN) feedback (see the discussion in \citealt{Pillepich_TNGmethods:2018}). Similarly, the galaxy size vs. halo size relation in original Illustris compared to IllustrisTNG has a higher normalization and a different slope \citep{karmakar:2023}. 

Several studies have shown that in these modern cosmological hydrodynamic simulations, contrary to the classical picture described above, the predicted galaxy size does not strongly correlate with the halo spin parameter \citep{Zjupa:2017,Teklu:2015,karmakar:2023}. \citet{Jiang:2019} carried out a study of two sets of high resolution ``zoom-in'' hydrodynamic simulations (NIHAO and VELA), confirmed the lack of correlation between size and halo spin, and found a significant (anti)-correlation between galaxy size (or $r_{\rm gal}/R_{\rm halo}$) and the Navarro-Frenk-White \citep{navarro:1997} \emph{concentration parameter} of the halo, $c_{\rm NFW}$. 

Another probe of how halo properties relate to galaxy sizes can be derived from galaxy clustering. \citet{Hearin:2019} split galaxies from the Sloan Digital Sky Survey (SDSS) into those with above-median sizes and below-median sizes in a specific stellar mass bin and compared their relative clustering. They found that, in a given stellar mass bin and at a given spatial scale, small galaxies cluster more strongly than large galaxies. It is well known that ``secondary'' halo properties (properties other than mass) tend to correlate with large scale environment and hence with the clustering strength (this gives rise to the effect commonly known as ``halo assembly bias'', which is defined as a dependence of clustering on halo properties other than mass). Different halo properties (e.g. spin, concentration, mass assembly history) show different characteristic clustering signatures \citep[e.g.][]{bose:2019,Hadzhiyska:2020,Hadzhiyska:2021,contreras:2021}. \citet{Behroozi:2022} considered several empirical models in which galaxy size is assumed to correlate with different secondary halo properties, including spin, concentration, and halo growth rate. They find that the spin and concentration based models are both inconsistent with the observed relative clustering of small vs. large galaxies. The model based on the halo growth rate averaged over the past dynamical time provides the best agreement with the measured size-dependent clustering. Their results are in direct tension with the correlation between galaxy size and halo concentration reported by \citet{Jiang:2019}, as well as with the spin-based models commonly adopted in semi-analytic models of galaxy formation. 

In this work, we make use of the IllustrisTNG suite of cosmological hydrodynamic simulations \citep{Pillepich_TNGmethods:2018,Nelson:2018} to investigate this puzzle. The IllustrisTNG simulations have been shown to predict galaxy size-mass relations for both star forming and quiescent galaxies that are in very good agreement with observations from $z\sim 0$--3 \citep{Genel:2018}. We have computed the properties of dark matter halos in the dark matter only N-body runs of IllustrisTNG and matched these halos to those in the full physics simulations \citep{Gabrielpillai:2021}. We directly investigate the correlations between galaxy size or $r_{\rm gal}/R_{\rm halo}$ and the halo spin, concentration, and formation history in the simulations. We note that in the past, most such studies have used halo properties extracted from the simulations that include baryons and baryonic physics, and are thus subject to ``contamination'' by baryonic effects. We then take advantage of the large volume TNG300 box to directly compute the relative clustering of small versus large galaxies --- the first time that this comparison has been done in a hydrodynamic simulation. We compare these results directly with both the observational results and the empirical models presented in \citet{Behroozi:2022}. Finally, we present predictions for the size-dependent clustering of galaxies at high redshift, which will be measurable with future high-resolution wide-field surveys such as those that will be carried out with the Nancy Grace Roman Telescope. 

The structure of this paper is as follows. In Section~\ref{sec:methods}, we briefly describe the IllustrisTNG simulation suite and how we measured halo properties from the matched dark matter only simulations. We also provide a brief summary of the observational results on differential clustering of galaxies as a function of their size.  In Section~\ref{sec:results}, we present our results. In Section~\ref{sec:discussion}, we discuss the implications of our results and the caveats of our analysis. In Section~\ref{sec:conclusions}, we summarize our main conclusions.

\section{Simulations, Methods, and Observational Data}
\label{sec:methods}
\subsection{Hydrodynamic simulations}
The IllustrisTNG suite of cosmological magneto-hydrodynamic simulations \citep{Pillepich_TNGmethods:2018,Springel:2018,Nelson:2018,Naiman:2018,Marinacci:2018} is based on the moving mesh hydrodynamic solver AREPO \citep{springel:2010}, and is the successors to the Illustris suite of simulations \citep{Vogelsberger:2014, Vogelsberger:2014b}. As with all large volume cosmological simulations, IllustrisTNG treats processes such as star formation, stellar feedback, metal enrichment, black hole seeding, black hole accretion, and AGN feedback using ``sub-grid'' recipes, as they cannot be resolved directly. The star formation prescription is based on the \citet{SH:2003} model and details are described in \citet{Vogelsberger:2014}. The implementation of stellar driven winds and other processes is described in \citet{Pillepich_TNGmethods:2018}. Details of the AGN feedback prescription, which was significantly modified in TNG relative to the original Illustris simulations, are given in \citet{Weinberger:2017}. 

The TNG suite is comprised of several runs with different box sizes and resolutions \citep{TNGdatarelease}. Each ``full physics'' run has a matched dark matter only run with the same initial conditions. We summarize the runs used in this work in Table ~\ref{tab:tng_table}. For additional details and results from the TNG50 runs, please see \citet{Nelson:2019} and \citet{Pillepich:2019}. 

\begin{table}
	\centering
	\begin{tabular}{lrrrr} 
		\hline
		Simulation & $L_{\rm Box} [{\rm cMpc}]$ & $N_{\rm DM}$ & $m_{\rm DM} [\msun]$  & $m_{\rm gas} [\msun] $\\
		\hline
		TNG50-1       & 51.7  & $2160^{3}$ &  $4.5 \times 10^5$ & $8.5 \times 10^4$ \\ 
		TNG50-1-Dark  & 51.7  & $2160^{3}$ & $5.4 \times 10^5$ & 0 \\ 
		TNG100-1      & 110.7  & $1820^{3}$ & $7.5 \times 10^{6}$ & $1.4 \times 10^6 $ \\
		TNG100-1-Dark & 110.7  & $1820^{3}$ & $8.9 \times 10^{6}$ & 0 \\
		TNG300-1      & 302.6 & $2500^{3}$ & $5.9 \times 10^{7}$ & $1.1 \times 10^7$ \\
		TNG300-1-Dark & 302.6 & $2500^{3}$ & $7.0 \times 10^{7}$ & 0 \\
		\hline
	\end{tabular}
    \caption{Summary of TNG runs used in this study, from \url{https://www.tng-project.org/data/}. $L_{\rm Box}$ is the box side length, $N_{\rm DM}$ is the number of dark matter particles, and $m_{\rm DM}$ is the mass of a dark matter particle, and $m_{\rm gas}$ is the gas mass resolution. 
    } 
	\label{tab:tng_table}
\end{table}

Halos and subhalos are identified in TNG using friends-of-friends (FoF) and the \subfind\ algorithm \citep{Springel:2001}. We adopt the stellar mass definition that is the sum of all stellar particles within the stellar half mass radius (\texttt{SubhaloHalfmassRadType}, \texttt{Type=4}). 
Indexing the FoF catalog via the \subfind\ catalog is done using \texttt{SubhaloGrNr}, the index of the parent FoF halo. \texttt{GroupFirstSub} identifies the most massive bound subhalo in the halo, which we call a central galaxy. Note that in this work, all halo properties are extracted from the dark-matter-only simulations using the \rockstar\ halo finder, as described below. For a comparison between \rockstar\ (DM only) and \subfind\ (full physics) halo mass estimates for TNG100, see \citet{Gabrielpillai:2021}. A summary of the definitions of all quantities used in this paper is provided in Table~\ref{tab:quantities}.

The cosmological parameters adopted in the IllustrisTNG suite are $\Omega_{m}$ = 0.3089, $\Omega_{\Lambda}$ = 0.6911, $\Omega_{b}$ = 0.0486, and $h$ = 0.6774. We note that these are nearly identical to the parameters adopted by B22.

\subsection{Measuring dark matter halo properties}
\label{subsection:dmprops}
In order to obtain robust measurements of the properties of the dark matter halos in the TNG suite, we have run the \rockstar\ halo finder \citep{rockstar} on the dark matter only runs specified in Table~\ref{tab:tng_table}. For more details about the \rockstar\ runs please see Section~3.1.1 of \citet{Gabrielpillai:2021}.  \rockstar\ computes halo properties within a spherical region with a specified overdensity. Many different conventions for the value of this overdensity are used in the literature. As in \citet{Gabrielpillai:2021}, we adopt the halo virial mass definition of \citet{Bryan:1998}. \rockstar\ also carries out a fit to the spherically averaged halo density profile to estimate the NFW concentration parameter, and computes the angular momentum and the dimensionless spin parameter of each halo. In this work, we adopt the \citet{Peebles:1969} definition of the spin parameter. 

The halo formation time, $z_{\rm form}$ is often defined as the redshift where the halo has assembled half of its mass, mathematically expressed as 
\begin{equation} 
\label{eq:z_form}
M_{\rm vir}(z_{\rm form}) = \frac{1}{2} M_{\rm vir}(z = 0) 
 \end{equation} 
 We first extract the most massive progenitor for each $z = 0$ root halo with at least 1000 DM particles. We also require that each progenitor halo contains at least 100 particles. 
 For each tree, we identify in which snapshot the halo mass history first exceeds 50\% of $m_{\rm vir}(z = 0)$, and record it and the previous snapshot. We then linearly interpolate the mass formation history and redshifts between the recorded snapshots to get  $z_{\rm form}$.   

As noted above, the publicly available TNG data products are based on halos and subhalos that were identified with a different halo finder, \subfind\ \citep{Springel:2001}, applied to the full physics simulations. We perform bijective matches between the \rockstar\ halos identified in the DM only runs and the \subfind\ ``Groups'' identified in the full physics runs. For more details, see \citet{Gabrielpillai:2021}. This allows us to assign properties identified via \rockstar, such as dimensionless halo spin, formation time, and concentration, to their bijectively matched \subfind\ halo. As shown in \citet{Gabrielpillai:2021}, it is more difficult to robustly identify the counterparts of satellite galaxies in the full physics TNG runs in the \rockstar\ catalogs based on the dark matter only runs. We therefore only use the bijectively matched sample for central galaxies. 

\subsection{Observational measurements of differential clustering}
In this work, we make use of the measurements of differential clustering of \citet{Behroozi:2022}.
We briefly summarize their analysis and results here, and refer to Section~4.2 of \citet{Behroozi:2022} for more details. 

The analysis is based on DR16 of the Sloan Digital Sky Survey \citep[SDSS;][]{Ahumada:2020}. Redshifts were taken from DR16, but stellar masses were taken from the \citet{Brinchmann:2004} MPA-JHU catalog, which is based on photometry from DR7. Stellar masses were converted to a Chabrier \citep{Chabrier:2003} stellar initial mass function.  Selection cuts are applied to obtain an approximately stellar mass complete sample, as described in \citet{Behroozi:2022}. 

B22 adopt the DR7 size measurements of \citet{Meert:2015}, which are based on fitting a two component Sersic plus exponential (SerExp) function to the projected light profile of each galaxy. The quantity that is more directly related to physical processes, and is generally measured in simulations, is the 3D radius of the stellar mass. B22 convert the observed quantities to estimated 3D stellar mass radii as described in their Section~4.2.2, assuming a fixed correction from light to mass as adopted in \citet{Somerville:2018}, and a projection correction based on the assumption that spheroidal galaxies are well represented by flattened \citet{Hernquist:1990} profiles. They note that the ratio of the clustering strength of large relative to small galaxies changes by less than 10\% when galaxies are divided by 2D half-light radii rather than 3D half-mass radii, which is well within the observational uncertainties.  

B22 then use the {\tt correl} utility distributed with {\sc UniverseMachine} to compute the projected redshift space angular correlation function $\xi(r_p, \pi)$. This is then integrated up to $\pi_{\rm max}=13.6$ h$^{-1}$ Mpc to obtain the projected correlation function $w_p(r_p)$.

\subsection{Measurements of differential clustering in simulations}
In the simulations, we compute the projected two-point correlation function $w_p(r)$ as follows:
\begin{equation}
    w_p(r) = \sum_{\pi=0}^{\pi_{\rm max}}\left[\frac{DD(r, \pi)}{RR(r, \pi)} - 1\right],
\end{equation}
where we are using the natural estimator due to the periodic boundary conditions of the box. Here, $\pi$ indicates the distance along the line-of-sight, and $DD(r, \pi)$ and $RR(r, \pi)$ are the normalized data-data and random-random pair counts as a function of pair distance in the perpendicular ($r$) and parallel ($\pi$) directions. The summation over $\pi$ indicates that we integrate pairs with line-of-sight separations out to $\pi_{\rm max} = 13$ h$^{-1}$ Mpc as in B22. The errorbar on each measurement is obtained by jackknifing. In particular, we split the box into 27 subboxes and excluding one subbox at a time, measure the projected clustering to obtain the mean and error on the two-point statistic. We perform this calculation for galaxies 
split into four stellar mass bins, ranging from $\log M = 9.75$ and 11.25, in units of $M_\odot$. 

\section{Results}
\label{sec:results}

\begin{figure*}
	\includegraphics[width=\textwidth]{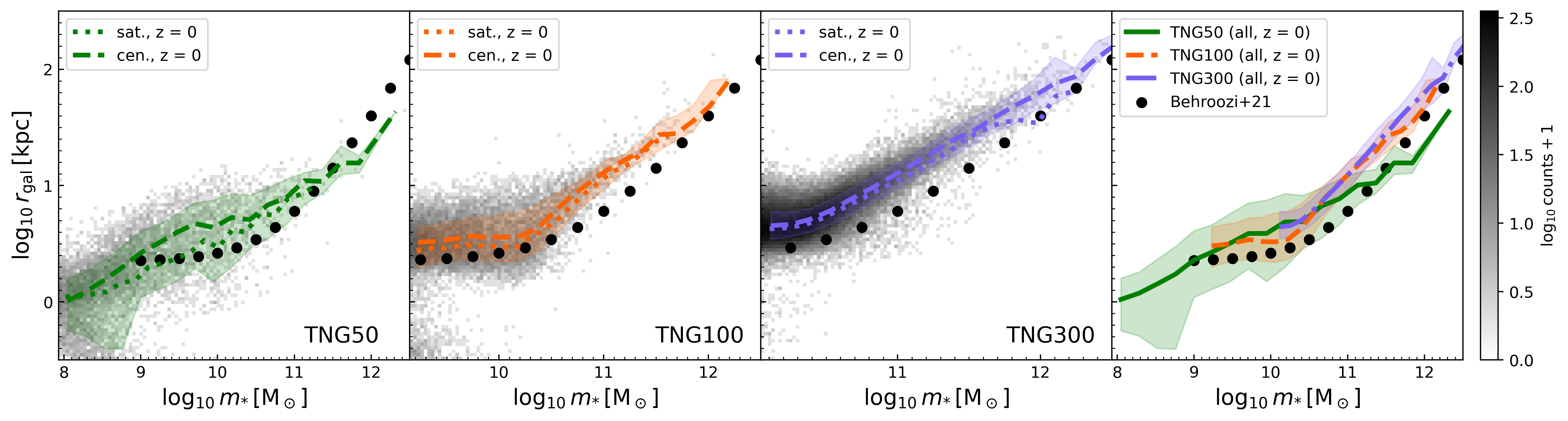}
    \caption{The size-mass relation for all three TNG boxes at z = 0. The first three panels from left to right show the median scaling relationship for central (dashed) and for satellite (dotted) and the 16-84 percentile region (as a shaded area) and individual galaxies (grey shaded regions) for TNG50 (green), TNG100 (orange), and TNG300 (purple). The fourth panel shows the median and 16-84 percentile areas for all three boxes. The size-mass relation for observed SDSS galaxies (Eq. 13 from \citealt{Behroozi:2022}) is overplotted on all four panels. The size-mass relation is similar for the three boxes, in spite of the differing resolution, and agrees qualitatively with the observations. } 
    \label{fig:sizemassz0}
\end{figure*}

\begin{figure*}
	\includegraphics[width=\textwidth]{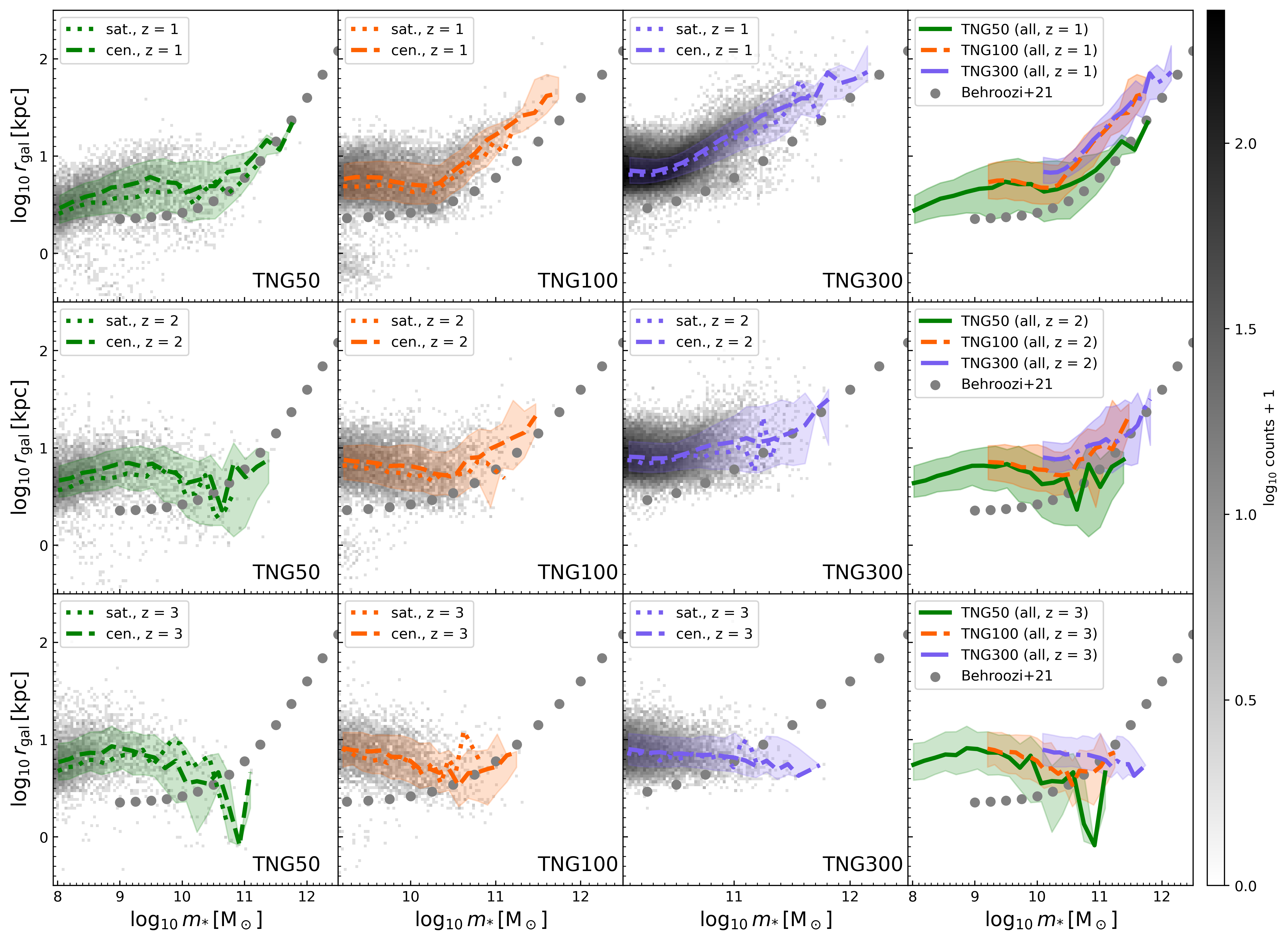}
    \caption{The size-mass relation for all three TNG boxes at (top) z = 1, (middle) z = 2, and (bottom) z = 3. Panels are as described in Fig. \ref{fig:sizemassz0}. The grey dots show the $z=0$ observed size-mass relation, and are only to guide the eye. } 
    \label{fig:sizemasshighz}
\end{figure*}

\subsection{The size-mass relation across the TNG suite}
In this sub-section we present the relationship between galaxy stellar mass and the 3D half-stellar mass radius in the TNG50, TNG100, and TNG300 volumes. In this sub-section, our sample consists of both central and satellite galaxies whose stellar mass is at least 100 stellar mass particles. 
Fig.~\ref{fig:sizemassz0} shows the median and 16th and 84th percentile spread of these relationships at $z=0$ for the three simulation volumes. It is well known that the stellar mass predictions for galaxies in TNG are not very well converged; i.e. the predicted stellar masses at a given halo mass depend on the numerical resolution of the simulation \citep{Pillepich_TNGmethods:2018}. In order to partially correct for this, we rescale the stellar masses in TNG300 by a factor of 1.4 at $z=0$, 1.3 at $z=1$, 1.1 at $z=2$ and do not rescale them at $z=3$, as suggested by \citet{Pillepich:2019}. Similarly, the convergence of the size predictions has not been explored extensively in TNG50, so we do not apply a correction. We do not attempt to correct the sizes or other properties for the effects of resolution, as it is unclear how to do so. 

We see fairly good agreement between the TNG50, TNG100, and TNG300 size predictions over the stellar mass range $9 \lesssim \log m_*/\msun \lesssim 10.5$, and more divergence at larger stellar masses $\log m_*/\msun \gtrsim 10.5$, with TNG50 producing the most compact galaxies at a given stellar mass, TNG300 producing the most extended galaxies, and TNG100 in between. We see qualitative agreement between the TNG100 predictions, and the observationally derived relation from B22, although the predicted sizes in TNG are about 0.2-0.3 dex higher than the B22 size estimates at $\log m_*/\msun \gtrsim 9.5$. A comparison between the observed size-mass relation for TNG100 has been shown previously by \citet{Genel:2018}, and shows a similar level of agreement, although the comparison there was done for star forming and quiescent galaxies separately, and was done for 2D projected light-based radii rather than the 3D mass-based radii used here. It is interesting that TNG100 shows a flattening in the stellar mass vs. size relation at low masses, while TNG50 predicts a monotonically decreasing size-mass relation at $z=0$ down to the lowest-mass resolved galaxies. 

Fig.~\ref{fig:sizemasshighz} shows the same quantity (stellar mass vs. 3D half-stellar mass radius) for the three TNG volumes at different redshifts, $z=1$--3. We note that the $z=0$ observational relation from B22 is overplotted on all panels to guide the eye, but the true observed relation is known to evolve somewhat with redshift \citet[e.g.][]{vanderwel:2014}. Due to the uncertainties in converting from light to mass based quantities and correcting for projection \citep[see the discussion in][]{Somerville:2018}, we do not carry out a detailed comparison with observations here. Once again, \citet{Genel:2018} compared the projected 2D half-light radii from TNG100 with the observations of \citet{vanderwel:2014} at $z=1$ and $z=2$ for star forming and quiescent galaxies, and found good agreement. It is interesting that by $z=1$, all three volumes show a flattening in the size-mass relation on the low-mass end (though most pronounced in TNG50 and TNG100), and by $z=3$, the relation shows an inversion, with low-mass galaxies actually having larger radii than massive galaxies in all three volumes. Once again, the three volumes show a lack of convergence on the high mass end, with TNG50 always predicting the most compact galaxies and TNG300 the most extended, with TNG100 in between. 

\begin{figure*}
	\includegraphics[width=\textwidth]{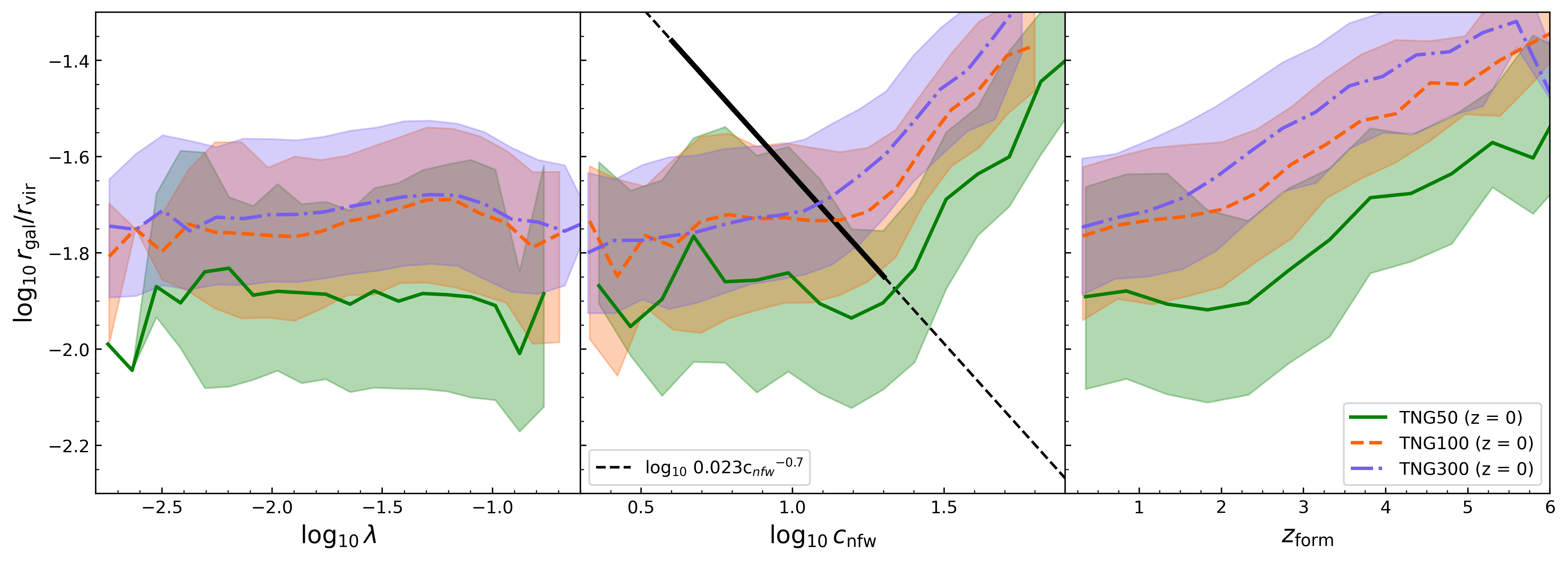}
    \caption{Ratio of galaxy radius to halo radius, $r_{\rm gal}/R_{\rm vir}$ vs. (left) spin, (middle) halo concentration, and (right) halo formation time at $z=0$ for TNG50 (green), TNG100 (orange), and TNG300 (purple). The black line in the middle panel shows the fit to the correlation reported by \citet{Jiang:2019}, where the thicker solid line indicates the approximate range of concentrations represented in the simulations used in that study. We find a negligible correlation between $r_{\rm gal}/R_{\rm vir}$ and spin, and a weak positive correlation between $r_{\rm gal}/R_{\rm vir}$ and halo concentration and formation time. Our results for $r_{\rm gal}/R_{\rm vir}$ vs. concentration are very different from those reported by \citet{Jiang:2019} based on a different suite of simulations (see text for discussion). } 
    \label{fig:halocorrz0}
\end{figure*}

\begin{figure}
	\includegraphics[width=\columnwidth]{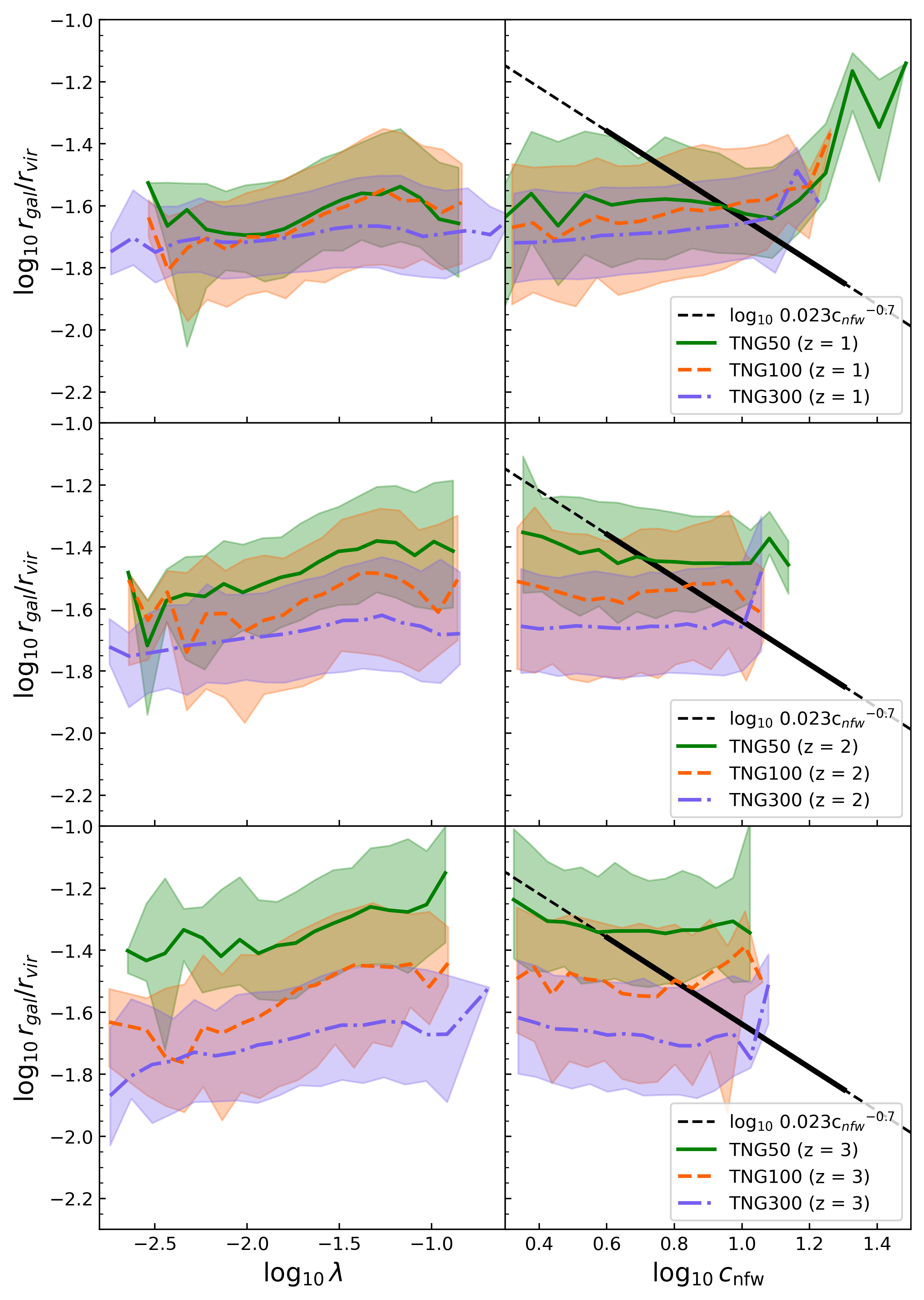}
    \caption{Ratio of galaxy radius to halo radius $r_{\rm gal}/R_{\rm vir}$ vs. (left) spin and (right) concentration for (top) z = 1, (middle) z = 2, and (bottom) z = 3. See Fig.~\ref{fig:halocorrz0} for key. The correlation between $r_{\rm gal}/R_{\rm vir}$ and halo spin becomes slightly stronger at high redshift, while the correlation between $r_{\rm gal}/R_{\rm vir}$ and halo concentration becomes weaker. }. 
    \label{fig:halocorrhighz}
\end{figure}

\subsection{Correlations between galaxy size and dark matter halo properties}
In this sub-section, we show how the ratio between galaxy 3D stellar half-mass radius and halo virial radius ($r_{\rm gal}/R_{\rm vir}$) is correlated with various dark matter halo properties, including the halo spin $\lambda$, the halo NFW concentration $c_{\rm NFW}$, and the halo formation redshift $z_{\rm form}$. In this section, our sample
consists of bijectively matched central galaxies where $m_{*} \geq 100 m_{\rm baryon}$ and $m_{\rm vir} \geq 1000 m_{\rm DM}$ for robust measurements of halo properties. The halo formation time is defined as the redshift when 50 percent of the final halo mass was first in a single progenitor (see Section~\ref{subsection:dmprops}). We note that importantly, we measure the halo properties on the \emph{dark matter only} simulations, and use the bijective matching technique described in Section~\ref{sec:methods} to link these halos to the appropriate halos in the full physics runs. Fig.~\ref{fig:halocorrz0} shows these correlations at $z=0$. As in many previous studies, we find no significant correlation between halo spin $\lambda$ and $r_{\rm gal}/R_{\rm vir}$. We also find no significant correlation with NFW concentration $c_{\rm NFW}$ at lower correlations $\log c_{\rm NFW} \lesssim 1.2$, and a significant positive correlation at $\log c_{\rm NFW} \gtrsim 1.2$. This behavior is qualitatively similar in all three volumes, although $r_{\rm gal}/R_{\rm vir}$ is in general offset to slightly lower values in TNG300, and TNG50 shows a pronounced dip in this relationship at $\log c_{\rm NFW} \sim 1.2$-1.4. Our results are quite strikingly different from those of \citet{Jiang:2019}, who found a \emph{negative} correlation between $r_{\rm gal}/R_{\rm vir}$ and $c_{\rm NFW}$. We discuss possible reasons for this discrepancy in Section~\ref{sec:discussion}. Perhaps the strongest correlation is seen between $r_{\rm gal}/R_{\rm vir}$ and formation redshift, such that earlier forming halos host more extended galaxies. This trend is again qualitatively similar in all three volumes. 


Fig.~\ref{fig:halocorrhighz} shows the correlations between $r_{\rm gal}/R_{\rm vir}$ and halo spin and concentration at $z=1-3$. It is well known that concentration and formation time/redshift are strongly correlated, so we did not deem it necessary to show both. Interestingly, we find that the correlation between $r_{\rm gal}/R_{\rm vir}$ and spin becomes slightly stronger with increasing redshift, especially in the TNG50 and TNG100 boxes. 
The discrepancy between the predicted values of $r_{\rm gal}/R_{\rm vir}$ among the different volumes also becomes larger towards higher redshift, both for the spin and the concentration. 

\begin{figure*}
	\includegraphics[width=\textwidth]{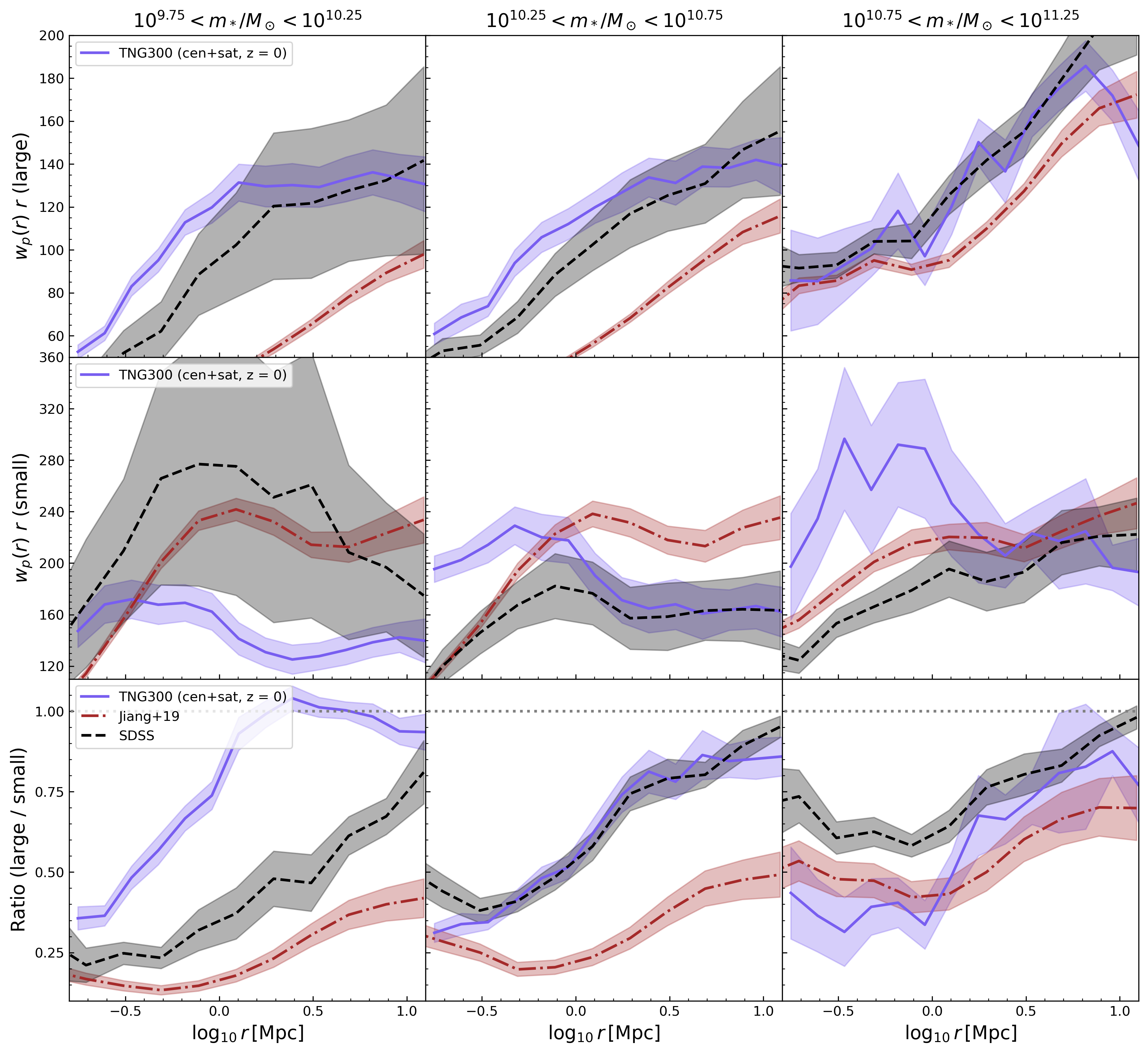}
    \caption{The 2-point correlation function as a function of projected distance for all galaxies (central and satellites combined) in TNG300 (purple), SDSS (grey), and the concentration based model from B22 (red) for three different stellar mass bins (three columns) at $z = 0$. We show the results for large (top) and small (middle) galaxies, as well as the ratio between the two (bottom). TNG shows scale-dependent size-based differential clustering, in qualitative agreement with the SDSS observations, but with quantitative differences, especially for small galaxies. The differential clustering in TNG is much weaker than what is predicted by the concentration based empirical model from B22. } 
    \label{fig:2ptallz0}
\end{figure*}

\begin{figure*}
	\includegraphics[width=\textwidth]{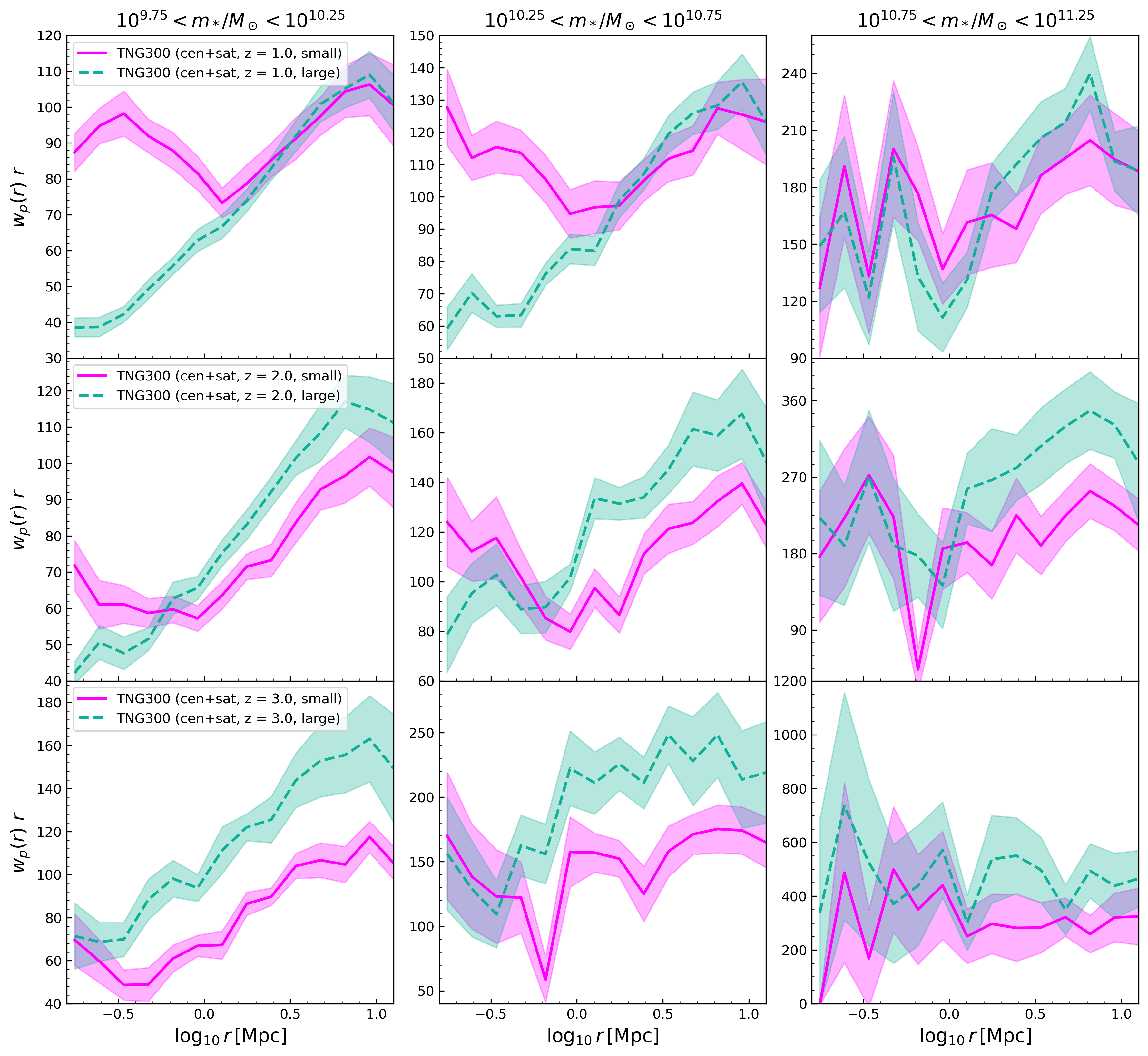}
    \caption{The 2-point correlation function for large (teal) and small (magenta) galaxies (central and satellites combined) for three different stellar mass bins at redshifts $z = 1$ (top), $z = 2$ (middle), and $z = 3$ (bottom), for the TNG300 simulation volume. } 
    \label{fig:2ptallhighz}
\end{figure*}

\begin{figure*}
	\includegraphics[width=\textwidth]{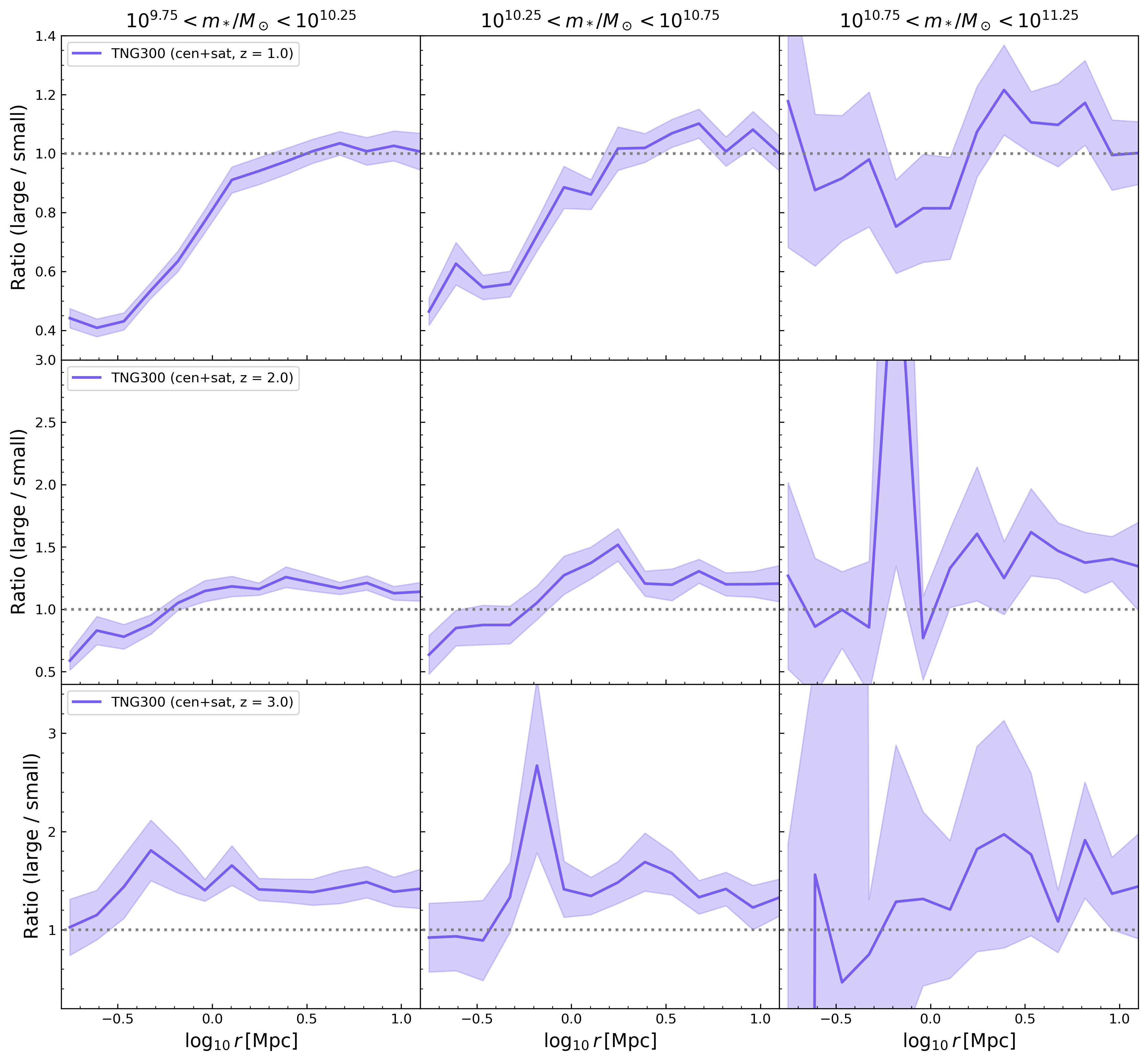}
    \caption{The ratio of the 2-point correlation functions for large vs. small galaxies, for central and satellite galaxies combined, for three different stellar mass bins at $z = 1$ (top), $z = 2$ (middle), and $z = 3$ (bottom). TNG predicts that size-based differential clustering flips in direction (large galaxies are less clustered than small ones at $z=0$--2 on small scales, but large galaxies become slightly more clustered than small ones at $z=3$). } 
    \label{fig:2ptratiohighz}
\end{figure*}

\subsection{Differential clustering as a function of galaxy size}
In this sub-section we show the projected two point correlation function $w_p$ in stellar mass bins, for ``large'' and ''small'' galaxies. Large galaxies are defined as having larger 3D half-stellar mass radii than the median in a given stellar mass bin, and small galaxies are similarly defined as having radii that are below the median. We only show results for the TNG300 box, as the other volumes are too small to obtain robust measures of clustering for sub-divided samples. Fig.~\ref{fig:2ptallz0} shows the two point correlation function $w_p$ as a function of projected distance $r$ at $z=0$. The columns show different bins in stellar mass, $9.75 < \log (m_*/\msun) < 10.25$, $10.25 < \log (m_*/\msun) < 10.75$, $10.75 < \log (m_*/\msun) < 11.25$, as adopted in B22. The top row shows  $w_p(r)$ for large galaxies, the middle row shows the same quantity for small galaxies, and the bottom row shows the ratio of $w_p(r_p)$ for large galaxies to that for small galaxies. We show the observational results from SDSS as computed by B22, along with the predictions from TNG300, and also the empirical model presented by B22 based on the correlation between galaxy size and halo concentration proposed by \citet{Jiang:2019}. We hereafter refer to this model as the B22 concentration based model. We see that the prediction of TNG300 is in very good agreement with the measurements from SDSS for large galaxies. The agreement is much less good for small galaxies, with small galaxies mostly being less clustered in TNG than they are in the observations in the lowest stellar mass bin, and more clustered in the higher stellar mass bins, especially at small separations. The scale dependence of $w_p(r)$ for small galaxies predicted in TNG is also quite different from the B22 observations. 

As already shown by B22, the concentration based empirical model makes predictions that are quite different from the observations, with a stronger differential clustering effect (a smaller ratio between the correlation functions of large and small galaxies). Perhaps unsurprisingly in view of the results of the previous sub-sections, we find that the concentration based model clustering predictions are also quite different from the measurements of differential clustering from the TNG300 full physics simulations. TNG300 always predicts a weaker differential clustering effect than the B22 concentration based model, at all scales and in all stellar mass bins (with the exception of scales $r_p \lesssim 1$ Mpc for the highest stellar mass bin, where the predicted ratio of the clustering of large vs. small galaxies in TNG300 is consistent with, or slightly lower than, the B22 concentration based model).  

Fig.~\ref{fig:2ptallhighz} shows the projected two point correlation function $w_p$ in the same stellar mass bins, for large and small galaxies, this time for redshifts $z=1$ (top), $z=2$ (middle), and $z=3$ (bottom). TNG300 shows an interesting trend in that in the two lowest stellar mass bins at $z=1$ and to a lesser extent at $z=2$, small galaxies show a strong upturn towards higher clustering values at small separation scales ($r \lesssim 1$ Mpc). This trend is not present at $z=0$, and has disappeared again by $z=3$. We delve into the physical origin of this trend in Section~\ref{sec:discussion}. 

Fig.~\ref{fig:2ptratiohighz} shows the ratio of the projected correlation function $w_p$ for large galaxies to that for small galaxies as a function of separation, again for $z=1$ (top), $z=2$ (middle), and $z=3$ (bottom). We note that in Fig.~\ref{fig:2ptallz0} we saw that at $z=0$, large galaxies in TNG were in general less clustered than small ones at a given stellar mass bin and separation (i.e. the ratio is less than unity), and there was a fairly strong trend in this ratio with separation, such that it approached unity at large separation scales ($r \sim 10$ Mpc). As we move to higher redshifts, we notice that the overall ratio moves closer to unity and even exceeds unity by $z=3$ (i.e., large galaxies are now \emph{more} clustered than small galaxies), and the trend with separation flattens out until it is almost completely absent by $z=3$.

\begin{figure*}
	\includegraphics[width=\textwidth]{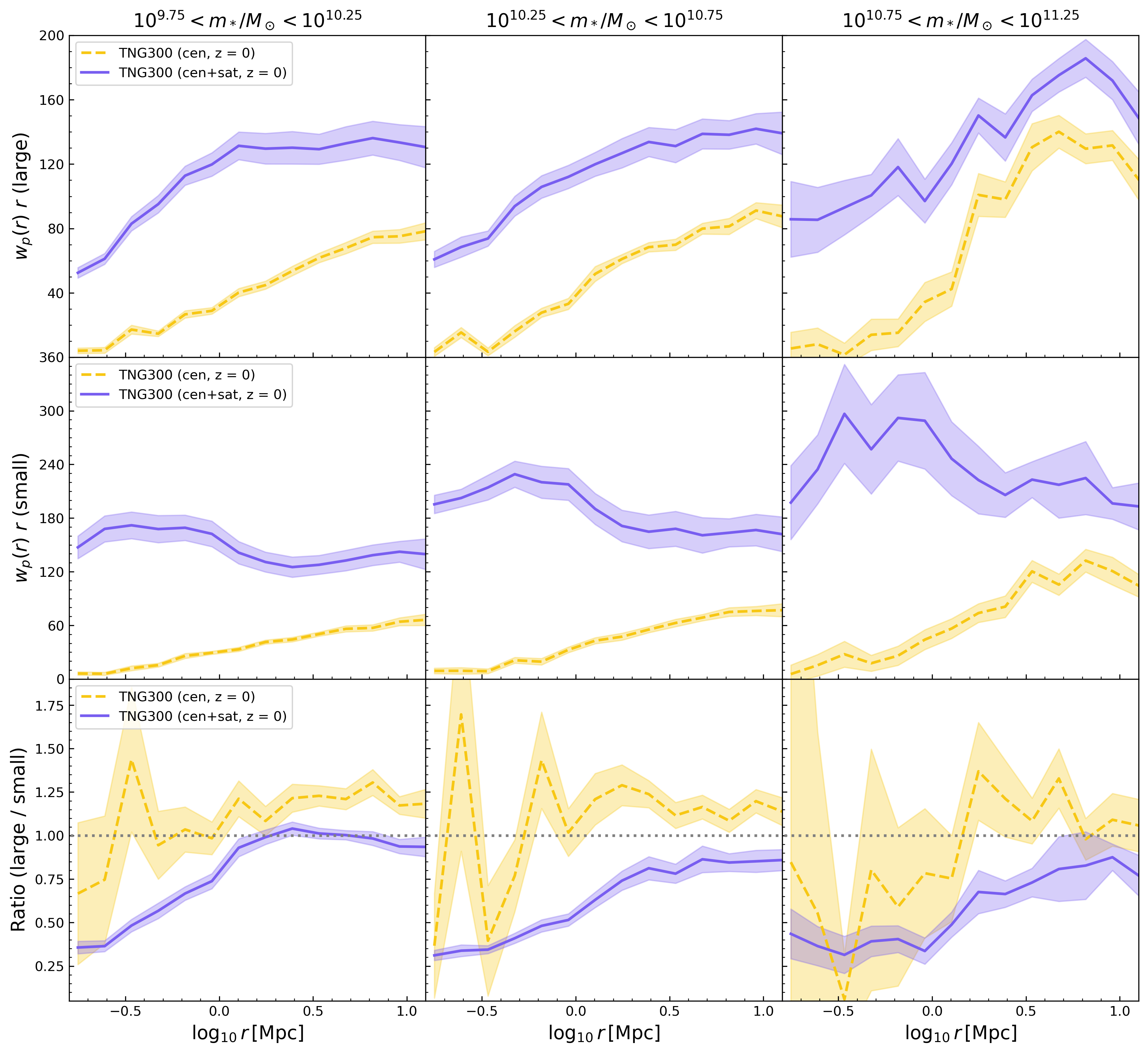}
    \caption{The 2-point correlation function as a function of projected distance for central galaxies only in TNG300 (purple)  at $z = 0$. We show the results for large (top) and small (middle) galaxies, as well as the ratio between the two (bottom). The plotted results are for centrals + satellite galaxies (purple) as shown in Fig. \ref{fig:2ptallz0} vs. centrals only (yellow). The size based differential clustering has become much weaker or largely disappeared, indicating that satellite galaxies play a crucial role in the differential clustering signal. }
    \label{fig:2ptcentralz0}
\end{figure*}
\citet{Hearin:2019} suggested that much of the differential clustering signal when galaxies are divided by size could be due to differing clustering properties and structural properties of satellite vs. central galaxies. We investigate this directly in Fig.~\ref{fig:2ptcentralz0}, where we divide galaxies into small and large as before (using the median size vs. radius relation for all galaxies) but then compute the clustering only for \emph{central} galaxies. It is striking that almost the entire differential clustering signal has now disappeared (the ratio of the clustering of large vs. small galaxies is close to unity), and although the measurement is quite noisy due to low numbers of galaxies, if anything large central galaxies are slightly \emph{more} strongly clustered than small central galaxies. 

\section{Discussion}
\label{sec:discussion}
In this section, we discuss the interpretation of our two main results --- the correlation of galaxy size with various halo properties and the predicted size-based differential clustering in TNG300 --- and how to reconcile the two. We discuss how our results compare with other related results in the literature, and review some of the caveats in the observational calculation as well as the TNG simulations. 

\begin{figure*}
	\includegraphics[width=\textwidth]{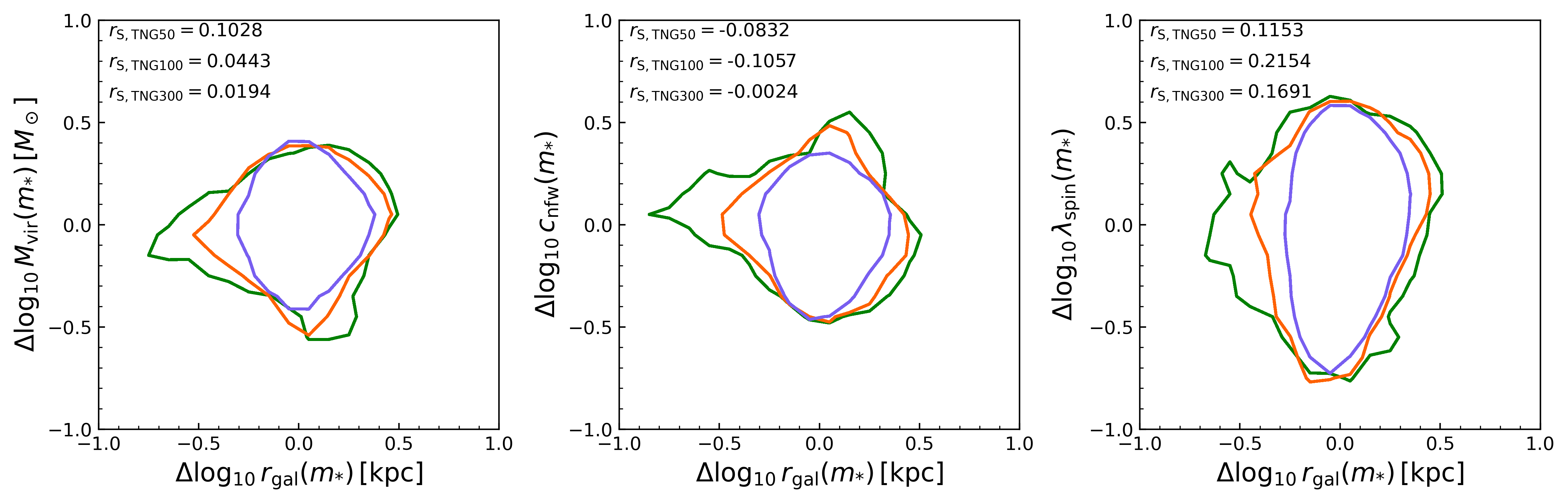}
    \caption{(Left) halo mass, (middle) halo concentration, and (right) halo spin residuals vs. galaxy size residuals for TNG50 (green), TNG100(orange), and TNG300 (purple). All residuals are functions of galaxy stellar mass. Annotated on each panel is Spearman's correlation coefficient for all three boxes.} 
    \label{fig:residuals-eagle}
\end{figure*}

\begin{figure*}
	\includegraphics[width=\textwidth]{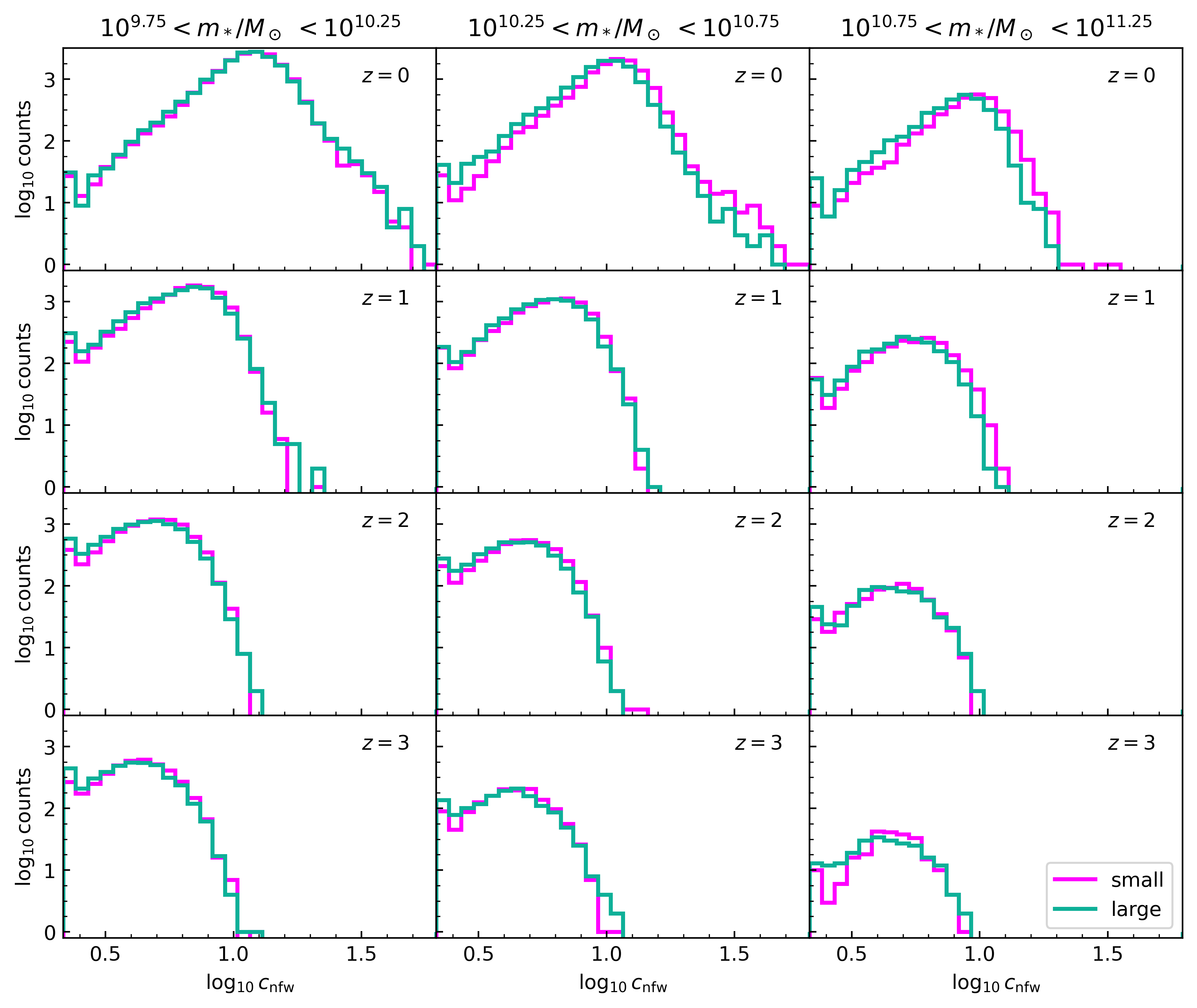}
    \caption{Distribution of the halo concentration $c_{\rm nfw}$ in the three stellar mass bins for TNG300 bijective matched central galaxies for z = 0, 1, 2, and 3, separated into populations of small and large galaxies. In the two lower stellar mass bins, the distributions are quite similar, while in the most massive bin, small galaxies live in systematically somewhat more concentrated halos at $z=0$--1. } 
    \label{fig:clustering-cnfw-dist}
\end{figure*}

\begin{figure*}
	\includegraphics[width=\textwidth]{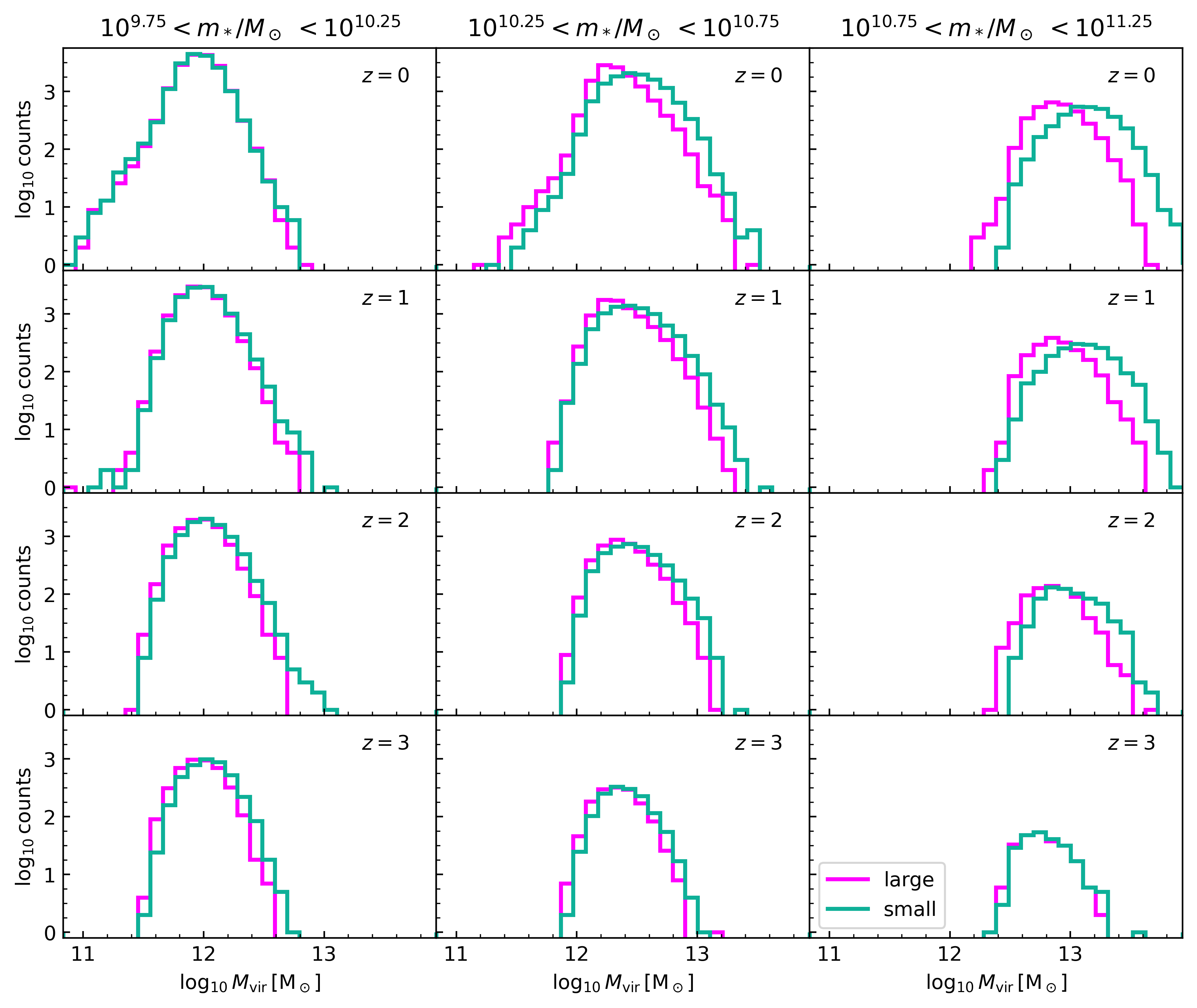}
    \caption{Distribution of $m_{\rm vir}$ in the three stellar mass bins for TNG300 bijective matched central galaxies for z = 0, 1, 2, and 3, separated into populations of small and large galaxies. Large galaxies live in systematically more massive halos in most stellar mass and redshift bins. This implies that $m_*/M_{\rm vir}$ depends on galaxy size. } 
    \label{fig:clustering-mvir-dist}
\end{figure*}

\subsection{The physical origin of galaxy sizes}
The goal of this work and many others is to answer the question: which physical processes are responsible for making some galaxies more extended and others more compact in radial size? The standard picture is that for rotation supported disks, the radial size is determined by the specific angular momentum of the gas that eventually forms the disk stars, and that this angular momentum arose from tidal torques during the collapse of the halo \citep{Blumenthal:1986,Dalcanton:1997,Mo:1998}. In this picture, assuming that there is no significant loss or transfer of specific angular momentum for the baryonic material during the disk formation process, we would expect a strong correlation between the spin parameter of the dark matter halo and the size of the galaxy \citep{Mo:1998}. However, the picture is complicated by the fact that galaxies also have diverse internal structure and kinematics. They differ in the radial functional form of their mass and light profiles, the ratio of vertical scale height to radius, and the fraction of support due to rotation vs. random motions. This structural, morphological, and kinematic diversity presumably reflects different physical formation channels. 

While the formation of disk galaxies was likely dominated by in-situ formation of stars fueled by relatively smooth gas accretion, spheroidal (or bulge-dominated) galaxies probably acquired much of their mass through mergers of other galaxies \citep{Oser:2010}. For a purely dissipationless (gas poor or ``dry'') merger, the radius of the remnant is larger than the radii of the progenitors \citep{Covington:2008,Covington:2011}. Thus, we might expect that for spheroid-dominated galaxies, the more gas-poor mergers a galaxy has experienced, the larger its size will be. However, if there is gas present in the merger progenitors, dissipation leads to more compact remnants \citep{Covington:2008,Covington:2011,Porter:2014}. 

Quite a few previous works have sought to gain insight into this question by studying correlations between galaxy size and dark matter halo properties in numerical hydrodynamic simulations, where all of the processes described above are explicitly modeled. For example, numerous works have studied whether the expected correlation between galaxy size and halo spin is present in such simulations \citep{Teklu:2015,zavala:2016,Zjupa:2017,karmakar:2023}. All such works have found a weak correlation between size and spin at best, with stronger correlations present when disk-dominated galaxies are selected \citep{yang:2023}.  \citet{Jiang:2019} also investigated the correlation of the galaxy size to halo size ratio $r_{\rm gal}/R_{\rm vir}$ with spin and concentration in two suites of zoom-in simulations (VELA and NIHAO). They again found no significant correlation with spin, but found a significant (anti-) correlation between $r_{\rm gas}/R_{\rm vir}$ and halo concentration, such that (smaller) more compact galaxies are found in more concentrated halos. 

We also investigate the correlation of $r_{\rm gal}/R_{\rm vir}$ with various halo properties. We are able to explore a larger dynamic range in halo properties than previous studies, as we combine together the three IllustrisTNG volumes. We find that halos less concentrated than $\log c_{\rm NFW} \lesssim 1.2$ show no significant correlation with concentration, or perhaps a weak anti-correlation (seen only in TNG50). However, for the first time we show that more concentrated halos show a strong \emph{positive} correlation between $r_{\rm gal}/R_{\rm vir}$ and concentration from $z\sim 0$--1. 

Both the VELA and NIHAO zoom-in simulation suites analyzed by J19 and the FIRE suite analyzed by \citet{el-badry:2016} span a limited range in halo mass ($10^{10}$--$10^{12} \msun$) and none of these simulations include AGN feedback. In simulations without AGN feedback, gas continues to accrete onto galaxies in massive halos, leading to ongoing active star formation, and overly compact sizes in well-known strong conflict with observations \citep{choi:2018}. While in the absence of AGN feedback, the trend of decreasing size with increasing concentration might be expected to continue, we suggest that quenching, driven by AGN, switches galaxies into a dry-merger dominated growth mode, which could counteract this trend. 

\citet{Desmond:2017} also analyzed a large volume cosmological simulation, EAGLE, which includes AGN feedback. They found a weak negative correlation (Spearman rank correlation coefficient of $-0.19$) between the residual in the size vs. stellar mass relation and the concentration vs. stellar mass relation. As we show in Fig.~\ref{fig:residuals-eagle}, although we observe a very weak correlation when using our full sample of galaxies in TNG100 (Spearman correlation coefficient $r=-0.11$), we find a more significant correlation ($r=-0.19$) when we restrict our sample to the stellar mass range used in \citet{Desmond:2017} ($ 9 < \log_{10} m_* / \msun < 11.5 $).
Binning together halos over a broader range of masses and concentrations may have washed out the correlation we identify only at high concentrations. Moreover, the EAGLE simulation used in their analysis has a volume of (100 Mpc)$^3$, and therefore does not provide the diversity or dynamic range of halo properties that our combined IllustrisTNG suite offers. 

We explicitly investigated the other halo properties of the host halos of the galaxies in the ``upturn'' part of the diagram of halo concentration and $r_{\rm gal}/R_{\rm vir}$ seen in the middle panel of Fig.~\ref{fig:halocorrz0} (i.e., the objects that are to the right of and above the black line). We find that these are predominantly \emph{low-mass} halos with unusually early formation times for their mass. Upon investigating these further, we find that their mass accretion histories are non-monotonic -- these halos increased in mass until some time in the past, whereupon the mass of their main progenitor began to decrease, presumably due to tidal stripping. We remind the reader that these are all central halos -- none-the-less, it is known that central halos can be stripped by the tidal field of nearby halos even when they are outside of the virial radius. It is also possible that these could be ``splashback'' halos (halos that entered the virial radius of another halo, then exited it again).  

In summary, our results suggest that correctly capturing the diverse channels of galaxy size growth (smooth gas accretion, gas rich and gas poor mergers) and their diverse environments is crucial in order to understand galaxy sizes and their dependence on halo properties. Moreover, it suggests that galaxy size may have a rather complex dependence on formation history, so that in order to understand the physical processes that determine galaxy size in hydrodynamic simulations, it is not enough to examine halo properties only a single output time. Methods that can ``learn'' correlations between the full merger history of a galaxy and global or structural properties, such as Graph Neural Networks (as used for example in the {\sc MANGROVE} method of \citealt{Jespersen:2022}), may be a promising avenue for future studies. 

\subsection{The physical origin of differential clustering}
One of the main results of this work is that the TNG300 simulation shows significant size-based differential clustering, in qualitative agreement with observations. To our knowledge, this is the first time that this has been demonstrated in a numerical cosmological hydrodynamical simulation. We can now attempt to determine the physical origin of differential size-based clustering in the simulation. 

Another important result is that Fig.~\ref{fig:2ptcentralz0} shows that nearly all of the differential clustering signal seen in TNG is due to satellite galaxies. When satellites are removed, the ratio of $w_p$ for large vs. small galaxies is close to unity within the large uncertainties due to small numbers of galaxies; if anything, large galaxies are slightly \emph{more} clustered than small galaxies, by up to about 20\%, in a reversal of the more significant trend seen when satellites are included. This lends very strong support to the interpretation put forward by \citet{Hearin:2019}:
they showed results based on dark matter only simulations, in which galaxy size was assumed to be proportional to halo radius at the redshift when the halo attained its maximum mass, with a fixed factor of proportionality. Satellite galaxies attain their maximum mass at an earlier time, when halos are more compact and their sizes are smaller, and therefore are assigned smaller sizes. This simple model, which includes \emph{no} additional dependence of galaxy size on secondary halo properties, produced remarkably good agreement with the observed size-based differential clustering measurements. 

Using the same projected two-point correlation function summary statistic considered here, \citet{Springel:2018} investigated the overall clustering of galaxies in the TNG100 and TNG300 volumes for stellar mass selected samples and for samples selected by both color and stellar mass, and compared their results with observational measurements of clustering from SDSS. They found very good overall agreement with the observational results for stellar mass selected samples, with clustering only slightly overestimated at small scales ($r_p \lesssim 1$ h$^{-1}$ Mpc), by at most perhaps 20\%.  They also find excellent qualitative agreement with the differential color-based clustering seen in SDSS, where red galaxies are much more strongly clustered at a given mass than blue galaxies. In two of the intermediate stellar mass bins ($9.0 < \log(m_*/$h$^{-2}\msun) < 9.5$ and $9.5 < \log(m_*/$h$^{-2}\msun) < 10$), the clustering of red galaxies is \emph{stronger} than that measured in the observations by up to a factor of 3--4. 

\citet{Hearin:2019} point out that in addition to being systematically smaller than central galaxies at fixed mass, satellites are expected to be systematically redder than central galaxies of the same mass. They also show that when galaxies are divided by color in addition to size and stellar mass, the differential size clustering signal in the observations almost disappears. Taken together, our results and those of \citet{Springel:2018} suggest that the treatment of satellite galaxies in IllustrisTNG may be inaccurate in some respects. This could be due to a number of factors, including the limited resolution of TNG300, which could lead to inaccuracy in the numerical treatment of tidal and ram pressure stripping, the sub-grid treatment of the ISM, or sub-grid modeling of stellar feedback. 

However, this leaves open the question of how to reconcile the observed clustering results with the dependence of galaxy size on secondary halo properties such as spin and concentration predicted by physics-based models and numerical simulations. B22 explored this using empirical models where galaxy sizes and stellar masses were assigned to halos based on different halo properties. They investigated four dicfferent scenarios:
\begin{itemize}
\item a \emph{spin-based} model, in which $r_{\rm gal} \propto \lambda R_{\rm vir}$, motivated by classical disk formation theory \citep{Mo:1998,Blumenthal:1986};
\item a \emph{concentration-based} model in which $r_{\rm gal} \propto c_{\rm NFW}^{-0.7} R_{\rm vir}$, motivated by the analysis of cosmological hydrodynamic zoom-in simulations by \citet{Jiang:2019};
\item a \emph{halo growth rate} model, in which galaxy size is correlated with the halo mass accretion rate $\dot{M}_{\rm vir}$ averaged over the past halo dynamical time, motivated by the analysis of cosmological hydrodynamic zoom-in simulations by \citep{el-badry:2016}; 
\item a ``null hypothesis'' model in which there is \emph{no} correlation between galaxy size and halo properties beyond stellar mass, motivated by the results of \citet{Desmond:2017} based on a large-volume cosmological hydrodynamic simulation.
\end{itemize}
They found that all spin-based models that they considered predicted a strong upturn in the clustering of large vs. small galaxies at small scales, which is not seen in the observations. Their concentration-based model predicted much \emph{stronger} size-based differential clustering than that seen in the observational data. The ``null hypothesis'' model produced negligible differential clustering, again in tension with the observational results. The only model that B22 found could reproduce the observed differential size-based clustering was the growth-rate based model. 

Are the B22 model and the \citet{Hearin:2019} model equivalent, or are they two alternative models that can both reproduce the observations? We note that \citet{Hearin:2019} emphasized that \emph{direct} environmental processes (such as tidal stripping of satellites) were \emph{not} the dominant driver of the differential clustering in their model; instead, the difference came down to smaller galaxies being hosted by earlier-forming halos, and satellites forming earlier than centrals of the same mass. As it is well known that the halo growth rate at an output time is tightly correlated with the halo formation time (halos that are rapidly growing today formed later than those that are growing slowly), the fundamental physical picture behind the \citet{Hearin:2019} and B22 models is essentially the same. 

We find that the clustering of large galaxies is in very good agreement with the observational measurement of $w_p$ from SDSS in all stellar mass bins considered (differing by at most $\sim$ 70\% on small scales). The predicted clustering of small galaxies in TNG is lower than that in SDSS (by up to a factor of two) on all scales in the lowest stellar mass bin, and is too high on scales $\lesssim 1$ Mpc in the two higher mass bins. This leads to a differential clustering signal ($w_p(\rm{large})/w_p(\rm{small})$) that is a bit too high in the lowest stellar mass bin, a bit too low on small scales in the highest stellar mass bin, and in (probably somewhat fortuitous) excellent agreement in the middle stellar mass bin. 

In order to summarize our resolution to the puzzle posed by B22, in Fig.~\ref{fig:clustering-cnfw-dist} and \ref{fig:clustering-mvir-dist} we show the distribution of host halo concentration and host halo mass for large and small central galaxies in the same stellar mass bins used in our clustering analysis (and that of B22). We can see that the distribution of $c_{\rm NFW}$ for small galaxies is \emph{very slightly} shifted to larger values, but the distribution of $M_{\rm vir}$ is shifted to larger values for \emph{large} galaxies, to a greater degree. It is relevant to consider there that it has been shown that in TNG there is a fairly strong correlation between halo concentration and $m_*/M_{\rm vir}$, such that more concentrated halos have larger $m_*/M_{\rm vir}$ \citep{Gabrielpillai:2021}. More massive halos cluster more strongly, but more concentrated halos at a fixed mass also cluster more strongly. These complex correlations between multiple galaxy and halo properties in high-dimensional space are not captured by empirical models such as those employed by B22. This highlights the power of such modeling approaches as an interpretive tool for more complex numerical hydrodynamic simulations. 

\subsection{Observational and Theoretical Caveats} 
In this sub-section we discuss sources of uncertainty in the observational measurements of size-based differential galaxy clustering, as well as in the IllustrisTNG numerical hydrodynamic simulations. 

On the observational side, surface brightness selection effects could impact the results by causing low surface brightness (extended) galaxies to be missing from the sample due to non-detection, while small galaxies that are too compact to be resolved could be mis-identified as stars. Galaxies that are very close to one another on the sky can be missed due to fibre collisions --- this is corrected for in the analysis of B22 by increasing the weights of galaxies that are closer than 55 arcsec on the sky. This analysis has been carried out in the space of stellar mass and 3D half-stellar mass radius, which must be computed from the observed fluxes and projected, light-based radii using a set of assumptions (for details, see B22). Both the correction from projected to 3D radius and from light-based to stellar-mass-based radius likely depend on galaxy shape, color, and color gradient, and the corrections applied by B22 do not fully account for these dependencies, which are complex and poorly understood. However, B22 state that the difference between light-to-mass corrections for elliptical vs. disk galaxies is small compared with the dispersion in the size-mass relation, and that their results differ by less than 10\% when using 2D vs. 3D radii. 

On the theory/simulation side, the largest uncertainties are due to what are commonly referred to as ``sub-grid'' recipes, i.e. phenomenological treatments of physical processes that occur below the resolution that can be explicitly simulated, including but not limited to star formation, stellar feedback, chemical enrichment, and black hole seeding, accretion, and feedback. IllustrisTNG is typical of state-of-the-art large volume cosmological hydrodynamic simulations in that many of these processes are treated using somewhat ad hoc recipes which have been calibrated to match a suite of observational constraints. It is well known that the sizes of galaxies predicted in numerical simulations are quite sensitive to the treatment of sub-grid processes such as stellar and AGN feedback, as well as potentially to the underlying hydro solver. Although IllustrisTNG reproduces a large number of key observational quantities, including some that it was not explicitly calibrated to match, there are some remaining tensions with observations which are presumably largely due to the limitations of the sub-grid modeling mentioned above. A related issue is the lack of convergence in both stellar mass and size between TNG simulations of different numerical resolution. Additionally, satellite galaxies in the lowest stellar mass bins used in our analysis may have too few particles, or the simulation may have too low a spatial resolution, to accurately simulate important environmental processes such as tidal and ram pressure stripping. 

\section{Conclusions}
\label{sec:conclusions}

At fixed stellar mass, some galaxies are observed to have much larger radii than others. The physical explanation for this dispersion in galaxy size, and the relationship between galaxy size and halo properties, is not well understood. In this study, we sought insight into this question by studying the emergent correlations between the ratio of galaxy to halo radius ($r_{\rm gal}/R_{\rm vir}$) and halo properties in the IllustrisTNG suite of cosmological hydrodynamical simulations. We extracted halo properties from dark matter only simulations with the same initial conditions, and matched them to galaxies in the full physics TNG simulations, for the three highest resolution runs of the TNG50, TNG100, and TNG300 volumes, yielding a large sample of halos over a broader range in mass than has previously been used for this type of study. We find the following main results: 
\begin{itemize}
\item After rescaling the stellar masses to partially correct for resolution effects, we find reasonable convergence in the predicted size-mass relation between TNG50, TNG100, and TNG300, except that TNG50 tends to predict more compact galaxies at high stellar mass ($m_* \gtrsim$ a few $10^{10}$--$10^{11} \msun$).
    \item In agreement with previous studies, we find a weak correlation between ($r_{\rm gal}/R_{\rm vir}$) and halo spin $\lambda$ at $z=0$. We find that this correlation was stronger at high redshift ($z\sim 2$--3).
    \item At $z=0$--1, we find no correlation between $r_{\rm gal}/R_{\rm vir}$ and halo concentration $c_{\rm NFW}$ at low to intermediate concentrations $c_{\rm NFW} \lesssim 16$, but a fairly strong positive correlation at higher concentrations. At $z\sim 2$--3, high concentration halos are missing from our sample, and there appears to be a somewhat stronger (though still weak) anti-correlation between $r_{\rm gal}/R_{\rm vir}$ and $c_{\rm NFW}$.
    \item We find a significant correlation between $r_{\rm gal}/R_{\rm vir}$ and the redshift at which the halo has assembled 50\% of its mass ($z_{\rm form}$). This is consistent with known strong correlations between $c_{\rm NFW}$ and $z_{\rm form}$.
    \item We find that the steeper slope of $r_{\rm gal}/R_{\rm vir}$ vs. $c_{\rm NFW}$ at high halo concentration is driven by halos that have had non-monotonic mass accretion histories; i.e., halos for which the peak mass of the largest progenitor occured in the past. 
\end{itemize}

Previous studies have highlighted the detection of a size-dependent differential clustering signal in observations from the SDSS survey \citep{Hearin:2019,Behroozi:2022}. Taking advantage of the relatively large volume of the TNG300 volume, we replicate this measurement on our simulated galaxy sample --- the first time, to our knowledge, that this calculation has been carried out with a cosmological hydrodynamic simulation. We divide galaxies into stellar mass bins, and within each mass bin, we use the measured median size in our simulated galaxy sample to further divide into small and large galaxy samples. We then proceed to compute the projected correlation function $w_p(r_p)$, following the same approach adopted by \citet{Behroozi:2022}. We find the following main results: 

\begin{itemize}
\item The TNG300 simulation exhibits significant size-based differential clustering in a sense that is qualitatively similar to the observational measurement, in the sense that large galaxies are \emph{less clustered} than small galaxies, and that the differential clustering signal is stronger at small scales and in lower stellar mass bins. 
\item There are significant quantitative disagreements between the TNG-based estimates of $w_p$ and the observational measurements for small galaxies, although the predicted $w_p(r_p)$ for large galaxies is in fairly good agreement with the observationally derived one. This leads to a differential clustering signal $w_p({\rm large})/w_p({\rm small})$ that is larger than the observationally derived one in the lowest stellar mass bin, and smaller in the highest stellar mass bin. 
\item We find that the differential size-based clustering signal becomes much weaker when we consider only central galaxies in TNG, strongly supporting the interpretation of \citet{Hearin:2019}, who proposed a model in which the galaxy size is proportional to the halo virial radius \emph{at the time when the halo reached its peak mass}. Satellite galaxies form earlier than central galaxies of the same mass, and halos were smaller in the past when the Universe was denser, so this leads to smaller sizes for satellite galaxies at a given stellar mass. 
\item In the two lower-mass bins, when satellite galaxies are removed, TNG300 predicts weak differential clustering in the opposite sense (large galaxies are more clustered).
\item The previous result is consistent with our findings, above, that central galaxies with larger radii tend to be hosted by halos of larger mass and higher concentration than small galaxies of the same stellar mass. 
\item We make predictions for size-based differential clustering at high redshift ($z=1$--3), finding that the differential clustering signal becomes weaker with increasing redshift out to $z\sim 2$, and then reverses in sense, so that $w_p({\rm large})/w_p({\rm small})>1$ at $z\sim 3$.
\end{itemize}

It will be interesting to follow up this work by exploring size-based differential clustering in future wide-area surveys with high resolution imaging, such as those that will be carried out with the Nancy Grace Roman Space Telescope.

\section*{Acknowledgements}

The Flatiron Institute is supported by the Simons Foundation. We gratefully acknowledge the use of Flatiron Institute computing facilities for this work. A.G. acknowledges that this material is based upon work supported by NASA under award number 80GSFC21M0002. B.H. thanks the Miller Institute for their generous support of her postdoctoral fellowship. 


\section*{Data Availability}

All data products used in this work will be made available upon reasonable request to the authors. 



\bibliographystyle{mnras}
\bibliography{sizeclustering}

\appendix
\section{Field definitions}
\label{app:fields}

Table \ref{tab:quantities} provides the definitions and provenances of quantities used in this work. We note that halo concentration ($c_{\rm NFW}$) is not recorded directly in the \rockstar\ catalog but can be computed from \texttt{Subhalo\_Rvir} ($R_{\rm vir}$) and \texttt{Subhalo\_Rscale} ($R_{\rm s}$), where $c_{\rm nfw} = R_{\rm vir} / R_{\rm s}$. The TNG database is available at \url{https://www.tng-project.org/data/}.

\begin{table*}
\begin{tabular}{lp{2.1cm}llp{5cm}}
\textbf{name in this paper} & source & \textbf{Name in catalog}            & \textbf{Native Units} & \textbf{Definition} \\ \hline
$m_*$ & TNG database & \texttt{SubhaloMassInHalfRadType} (\texttt{Type = 4}) & $10^{10} \ \msun / h$  & galaxy (subhalo) stellar mass within the stellar half mass radius   \\
$r_{\rm gal}$ & TNG database & \texttt{SubhaloHalfmassRadType} (\texttt{Type = 4})  & ${\rm ckpc} / h$  & (comoving) radius containing half of galaxy's (subhalo's) stellar mass \\
$\lambda$ & \rockstar\ catalog & \texttt{Subhalo\_Spin} & -- & Dimensionless halo spin parameter as defined by \citet{Peebles:1969} \\
$M_{\rm vir}$ & \rockstar\ catalog & \texttt{SubhaloMassType} (\texttt{Type = 1}) & $\msun / h$ & Halo virial mass as defined in \cite{Bryan:1998} \\
$R_{\rm vir}$ & \rockstar\ catalog & \texttt{Subhalo\_Rvir} & ${\rm cMpc} / h$ & Halo (comoving) virial radius as defined by \cite{Bryan:1998} \\
$c_{\rm NFW}$ & \rockstar\ catalog & See Appendix \ref{app:fields} & -- & Dimensionless halo concentration parameter as defined in \cite{navarro:1997} \\
$z_{\rm form}$ & \ct\ catalog & See Eq. \ref{eq:z_form} & -- & Redshift where a $z = 0$ halo's progenitor has assembled 50\% of its final mass \\
N/A & \rockstar\/-TNG bijective match catalog & \texttt{SubhaloIndexDarkRockstar\_SubLink} & -- & Bijective match indices between subhalos in the \rockstar\ and TNG catalogs as described in \cite{Gabrielpillai:2021} \\
N/A & TNG database & \texttt{GroupFirstSub}   & -- & Index of the FoF halo's most-massive (central) \subfind\ subhalo \\
N/A & TNG database & \texttt{SubhaloGrNr}   & -- & Index of a \subfind\ subhalo's parent FoF halo \\\hline
\end{tabular}
\caption{Table listing all quantities used in this work and their provenance.  }
\label{tab:quantities}
\end{table*}

\section{Supplemental Figures}
\begin{figure*}
	\includegraphics[width=\textwidth]{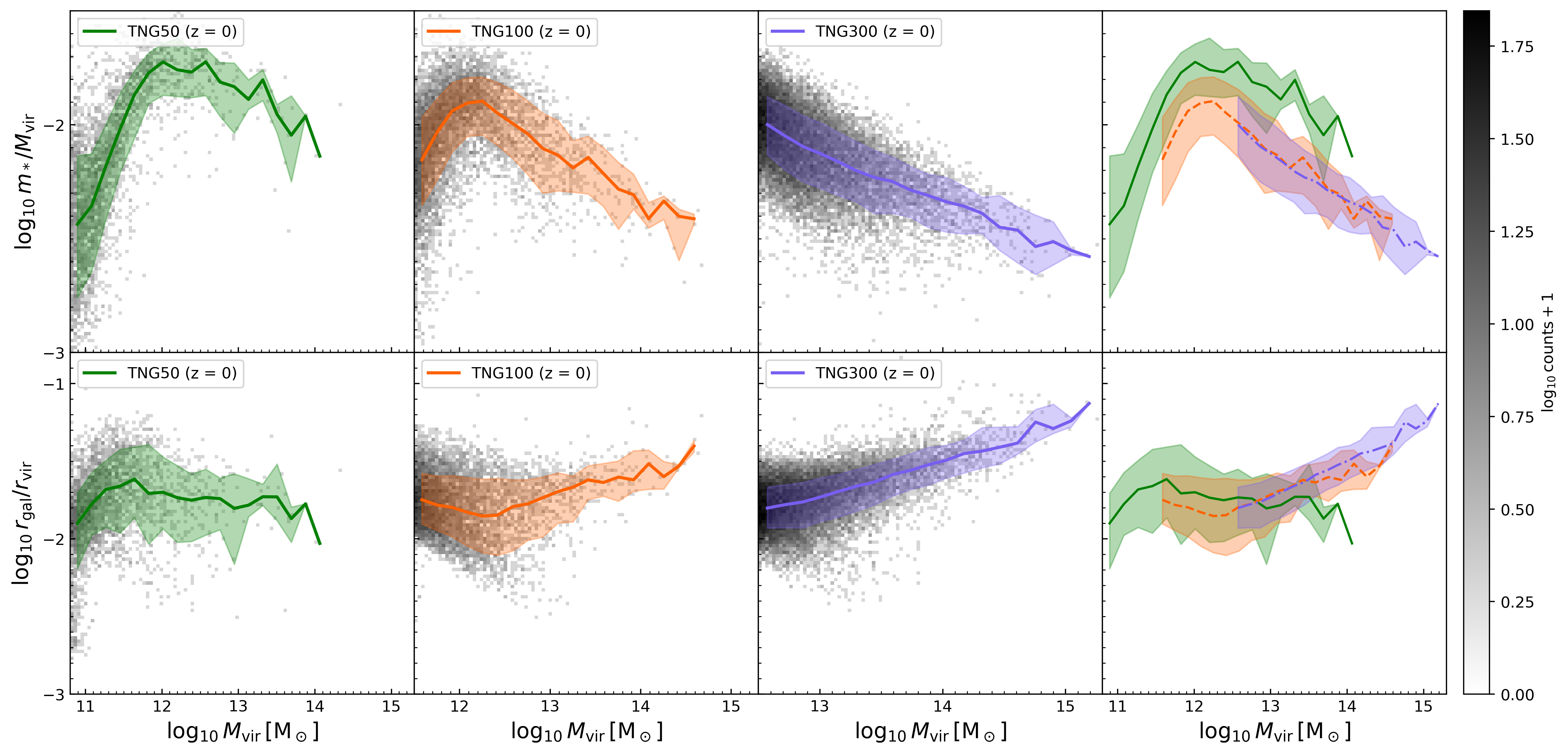}
    \caption{(top row) Stellar-to-halo mass relationship and (bottom row) $r_{\rm ratio}$ vs. $m_{\rm vir}$ scaling relationship for (from left to right): TNG50, TNG100, and TNG300. The median and shaded 16-84th percentile region is shown for all three boxes. Stellar masses in TNG300 have been rescaled as described in the main text. } 
    \label{fig:smhm-rel}
\end{figure*}

\begin{figure*}
	\includegraphics[width=\textwidth]{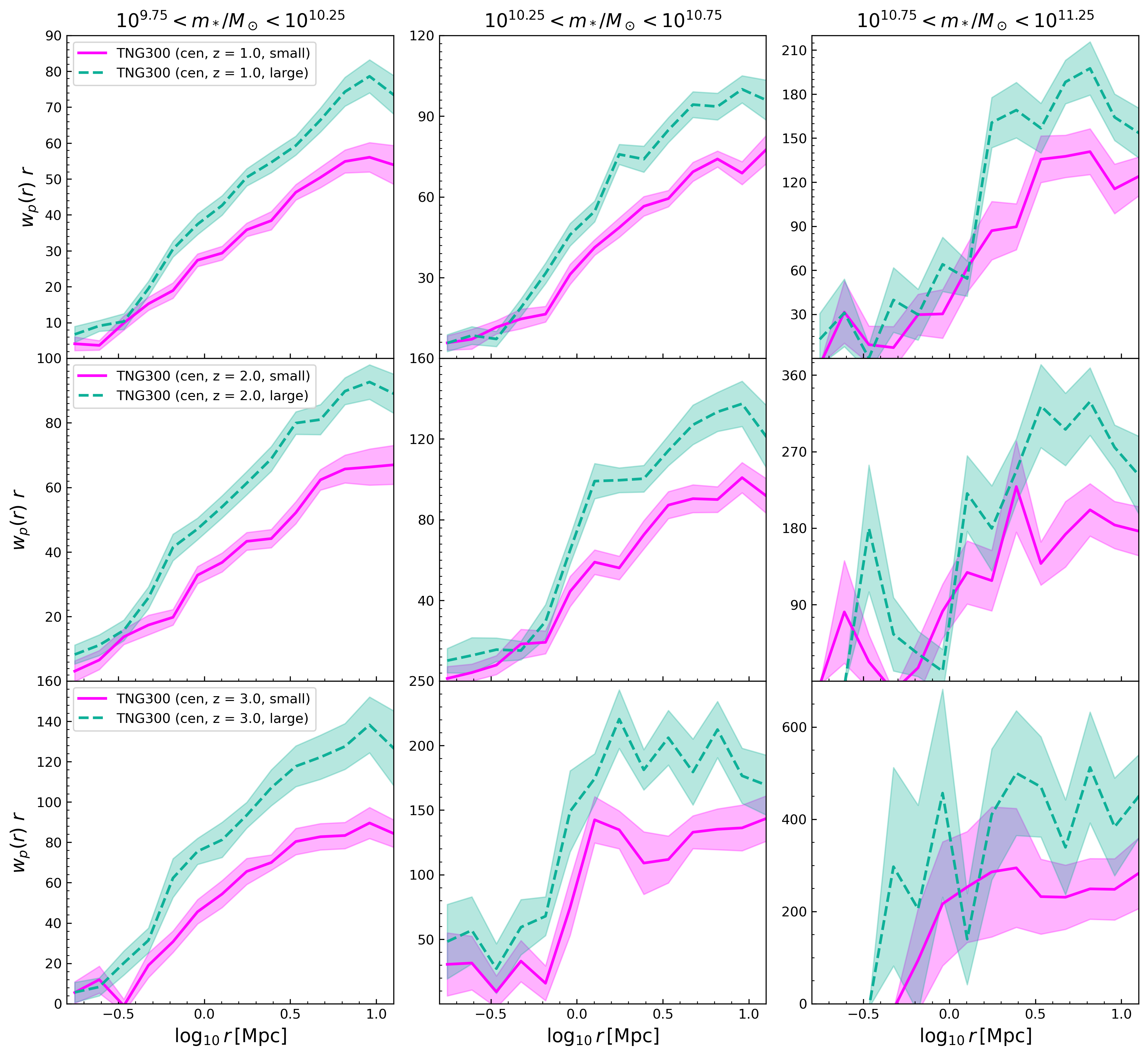}
    \caption{The 2-point correlation function for large (teal) and small (magenta) galaxies (central galaxies only) for three different stellar mass bins (three columns) at redshifts $z = 1$ (top), $z = 2$ (middle), and $z = 3$ (bottom), for the TNG300 simulation volume. } 
    \label{fig:2ptcentralhighz}
\end{figure*}

\begin{figure*}
	\includegraphics[width=\textwidth]{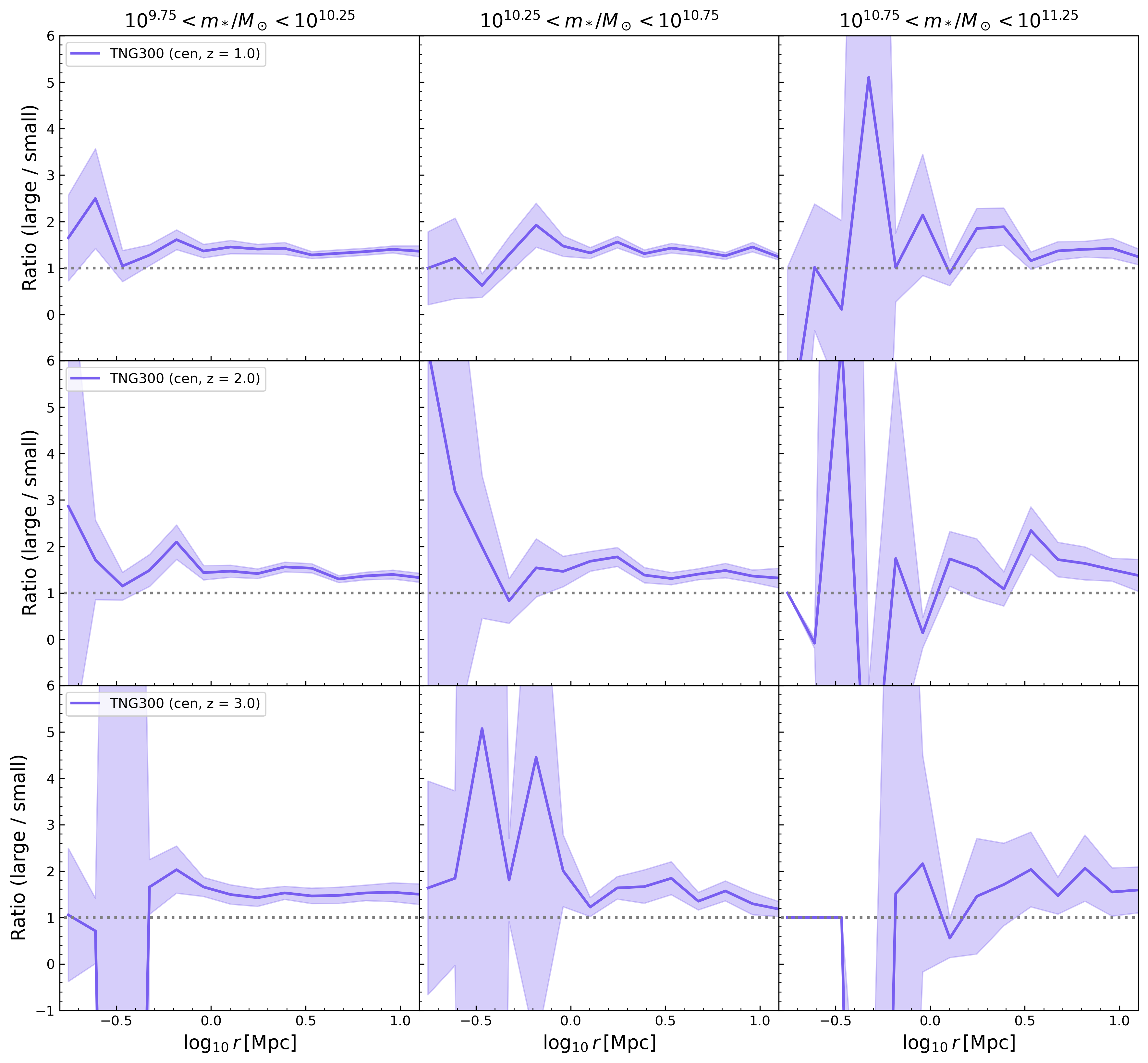}
    \caption{The ratio of the 2-point correlation functions for large vs. small galaxies, for central galaxies only, for three different stellar mass bins at $z = 1$ (top), $z = 2$ (middle), and $z = 3$ (bottom). } 
    \label{fig:2ptcentralratiohighz}
\end{figure*}

In this appendix, we show plots that are similar to ones in the main body of this work. Fig. \ref{fig:smhm-rel} shows both the stellar-to-halo mass relationship and galaxy size-halo mass scaling relationship for bijectively matched halos in all three boxes. Figs. \ref{fig:2ptcentralhighz} and  \ref{fig:2ptcentralratiohighz} are central-only analogs to Figs. \ref{fig:2ptallhighz} and \ref{fig:2ptratiohighz} respectively.

\bsp	
\label{lastpage}
\end{document}